\documentclass[nofootinbib,preprintnumbers,superscriptaddress,twocolumn]{revtex4}
\usepackage[dvipdf,dvips,dvipdfmx]{graphicx}
\usepackage{amsmath}
\usepackage{amssymb}
\usepackage{float}
\usepackage{wrapfig}
\usepackage{bm}
\usepackage{color}
\usepackage{comment}
\catcode`\@=11
\def\lsim{\mathrel{\mathpalette\@versim<}}
\def\gsim{\mathrel{\mathpalette\@versim>}}
\def\@versim#1#2{\vcenter{\offinterlineskip
\ialign{$\m@th#1\hfil##\hfil$\crcr#2\crcr\sim\crcr } }}
\catcode`\@=12

\newcommand{\p}{\partial}

\newcommand{\al}[1]{\begin{align}#1\end{align}}
\newcommand{\bp}{\begin{pmatrix}}
\newcommand{\ep}{\end{pmatrix}}
\newcommand{\nn}{\nonumber\\}
\newcommand{\paren}[1]{\left(#1\right)}
\newcommand{\sqbr}[1]{\left[#1\right]}

\newcommand{\eV}{\,\text{eV}}
\newcommand{\GeV}{\,\text{GeV}}

\newcommand{\df}{\text{d}}

\newcommand{\bs}[1]{\boldsymbol}

\newcommand{\Tr}{{\rm Tr}\,}

\newcommand{\fn}[1]{\!\left(#1\right)}
\newcommand{\cosmo}{{\tilde \Lambda}}
\newcommand{\newton}{{\tilde G}}
\newcommand{\mphi}{{\tilde m}_\phi}
\newcommand{\mchi}{{\tilde m}_\chi}
\newcommand{\lphi}{{\tilde \lambda}_\phi}
\newcommand{\lchi}{{\tilde \lambda}_\chi}
\newcommand{\lpchi}{{\tilde \lambda}_{\phi\chi}}

\usepackage{etoolbox}
\let\bbordermatrix\bordermatrix
\patchcmd{\bbordermatrix}{8.75}{4.75}{}{}
\patchcmd{\bbordermatrix}{\left(}{\left[}{}{}
\patchcmd{\bbordermatrix}{\right)}{\right]}{}{}

\allowdisplaybreaks

\begin{document}

\title{
Quantum gravity fluctuations flatten the Planck-scale Higgs potential
}

\author{Astrid \surname{Eichhorn}}
\email{a.eichhorn@thphys.uni-heidelberg.de}
\affiliation{Institut f\"ur Theoretische Physik, Universit\"at Heidelberg, Philosophenweg 16, 69120 Heidelberg, Germany}

\author{Yuta \surname{Hamada}}
\email{yhamada@wisc.edu}
\affiliation{Department of Physics, University of Wisconsin, Madison, WI 53706, USA}
\affiliation{KEK Theory Center, IPNS, KEK, Tsukuba, Ibaraki 305-0801, Japan}

\author{Johannes \surname{Lumma}}
\email{j.lumma@thphys.uni-heidelberg.de}
\affiliation{Institut f\"ur Theoretische Physik, Universit\"at Heidelberg, Philosophenweg 16, 69120 Heidelberg, Germany}

\author{Masatoshi \surname{Yamada}}
\email{m.yamada@thphys.uni-heidelberg.de}
\affiliation{Institut f\"ur Theoretische Physik, Universit\"at Heidelberg, Philosophenweg 16, 69120 Heidelberg, Germany}

\begin{abstract}
We investigate asymptotic safety of a toy model of a singlet-scalar extension  of the Higgs sector including two real scalar fields under the impact of quantum-gravity fluctuations.
Employing functional renormalization group techniques, we search for fixed points of the system which provide a tentative ultraviolet completion of the system.
We find that in a particular regime of the gravitational parameter space the canonically marginal and relevant couplings in the scalar sector---including the mass parameters---become irrelevant at the ultraviolet fixed point. 
The infrared potential for the two scalars that can be reached from that fixed point is fully predicted and features no free parameters. In the remainder of the gravitational parameter space, the values of the quartic couplings in our model are predicted.
In light of these results, we discuss whether the singlet-scalar could be a dark-matter candidate.
Furthermore, we highlight how ``classical scale invariance"  in the sense of a flat potential of the scalar sector at the Planck scale could arise as a consequence of asymptotic safety.
\end{abstract}
\maketitle 
\section{Introduction}
Compelling astrophysical and cosmological evidence points to the existence of dark matter. In its simplest form, dark matter might be just an additional scalar field. 
To stabilize the additional field and prevent it from decaying, a $\mathbb{Z}_2$ reflection symmetry can be used. Then, a dimension 4 operator exists that is compatible with the symmetries and couples the dark scalar to the standard model (SM) Higgs. The corresponding coupling is called  
Higgs portal coupling,
\cite{Silveira:1985rk,McDonald:1993ex,Bento:2000ah,Burgess:2000yq,Bento:2001yk,McDonald:2001vt,Davoudiasl:2004be,
Patt:2006fw, O'Connell:2006rsp,Barger:2007im,He:2007tt,Barger:2008jx,He:2008qm}.
The Higgs portal coupling is an additional marginal coupling and therefore expected to play an important role in a general effective field theory setup. It is particularly attractive, because in addition to providing a portal into the dark sector that enables direct and indirect experimental searches \cite{Cline:2013gha,Beniwal:2015sdl}, it could also contribute to stabilizing the Higgs potential
\cite{Clark:2009dc,Lerner:2009xg,Gonderinger:2012rd,Chen:2012faa,Gonderinger:2009jp,Khoze:2014xha,Eichhorn:2014qka,Hamada:2014xka,Hamada:2017sga}. 

Direct and indirect searches for a dark scalar have so far succeeded in constraining the allowed parameter space very significantly \cite{Cline:2013gha,Beniwal:2015sdl,Aprile:2015uzo,Akerib:2016vxi,Aprile:2017iyp,Cui:2017nnn}. One might thus wonder whether the dark sector is more complicated than just one extra scalar field, or whether there might be a fundamental reason why the dark scalar is ``hiding" from us. 
In this paper, we highlight that the asymptotic safety paradigm could provide a fundamental reason why the dark scalar has remained undetected.
Asymptotic safety is a generalization of asymptotic freedom, and provides a second alternative for a consistent microscopic regime of a quantum field theory: The running of couplings under the impact of quantum fluctuations can either lead into singularities---signaling a breakdown of the model---or into a scale invariant renormalization group (RG) fixed point regime.
For asymptotic freedom that fixed point is Gaussian, and therefore easily accessible by perturbative techniques. In the case of asymptotic safety, the fixed point is an interacting one.
Compelling hints for the existence of an asymptotically safe fixed point in gravity exist
\cite{Hawking:1979ig,Reuter:1996cp,Benedetti:2009rx,Christiansen:2012rx,Falls:2013bv,Christiansen:2014raa,Christiansen:2015rva,Gies:2016con,Christiansen:2016sjn,Denz:2016qks,Gonzalez-Martin:2017gza,Christiansen:2017bsy}; see also \cite{Niedermaier:2006wt,Niedermaier:2006ns,Percacci:2007sz,Reuter:2012id,Bonanno:2017pkg,Eichhorn:2017egq} 
for reviews. Within simple approximations, the fixed point persists under the impact of SM matter fields
\cite{Labus:2015ska,Oda:2015sma,Dona:2015tnf,Meibohm:2015twa,Hamada:2017rvn,Wetterich:2017ixo,Knorr:2017fus,Eichhorn:2017sok,Christiansen:2017cxa,Knorr:2017mhu} 
and asymptotically safe quantum fluctuations of matter impact the running of the SM couplings beyond the Planck scale \cite{Eichhorn:2011pc,Eichhorn:2012va,Meibohm:2016mkp,Eichhorn:2016esv,Eichhorn:2016vvy,Eichhorn:2017eht}. 
First hints suggest that a quantum-gravity induced ultraviolet (UV) completion for the SM might even allow to predict the Higgs mass \cite{Shaposhnikov:2009pv,Bezrukov:2012sa} and the top mass \cite{Eichhorn:2017ylw}, as well as the value of the Abelian gauge coupling \cite{Harst:2011zx,Christiansen:2017gtg,Eichhorn:2017lry,Eichhorn:2017muy}.
In our work, we find that under the impact of asymptotically safe quantum fluctuations of gravity, an UV completion of a toy model of the Higgs portal sector featuring two real scalar fields is induced in a simple approximation of the RG flow. In particular, the asymptotic safety paradigm appears to have a higher predictive power for the quartic couplings than a standard effective field theory setup, and in our approximation all quartic couplings---including the Higgs self-coupling, the dark scalar self-coupling and the Higgs portal coupling---are calculable quantities. Specifically, we find indications that quantum fluctuations of gravity force the Higgs portal coupling to vanish at and beyond the Planck scale. 
The underlying reason behind a flat potential for our toy model of an extended Higgs sector is shift symmetry, which appears to be protected under the impact of asymptotically safe quantum gravity \cite{Eichhorn:2017eht}.
Unlike the Higgs self-coupling, which is regenerated by SM fluctuations even if it is set to zero at the Planck scale, the Higgs portal coupling remains zero at all scales once it is set to zero at the Planck scale. Thus our scenario could provide an explanation for the lack of detections of scalar dark matter. We highlight that, e.g., the misalignment mechanism allows to produce all of the observed dark matter abundance. 

No sign of supersymmetry and other new physics at the LHC until now could suggest that the usual notion of naturalness might play a less central role in the identification of the scale of new physics than previously suspected.
 In this respect, an intriguing observation\footnote{The other interesting observation is that the Veltman condition~\cite{Veltman:1980mj} is satisfied at the Planck scale~\cite{Hamada:2012bp}, which might have some relevance with the naturalness problem~\cite{Chankowski:2014fva,Lewandowski:2017wov}.} was made by Bardeen~\cite{Bardeen:1995kv}:
Since the Higgs mass is the only dimensionful parameter in the SM, its tiny value at the Planck scale implies that the SM is close to being scale invariant.
That is, so-called classical scale invariance could be a naturalness condition for the SM.
It prohibits the existence of a mass scale in a classical action.\footnote{
In the literature, classical scale invariance is sometimes also called classical conformal invariance although the conformal symmetry is a larger group than the scale symmetry and the latter might not be sufficient to imply the former in QFTs in four dimensions. 
}
In a similar spirit, many models based on the classical scale invariance were proposed, recently. However in these models the underlying reasoning for scale invariance at the Planck scale is still lacking. In this paper, we show that the asymptotic safety paradigm could automatically generate a (nearly) scale invariant potential at the Planck scale.

\section{Method and Model}

\subsection{Functional renormalization group}
The functional renormalization group (FRG) is a method to evaluate the path integral by integrating the flow of the effective action under the momentum-shell-wise inclusion of quantum fluctuations. The central object in the FRG is the effective action $\Gamma_k$ with an infrared cutoff $k$, which contains the quantum corrections accumulated by integrating out the fluctuations with momenta $p^2>k^2$. The effective action is specified by a point in the theory space which is spanned by the couplings of an infinite number of effective operators respecting the symmetries of the model. Finite values for all couplings are generically generated by integrating quantum fluctuations, even if the effective action at some initial scale contains only a finite number of couplings.

The change of $\Gamma_k$ is described by a functional differential equation \cite{Wetterich:1992yh}  
\al{
\p_t \Gamma_k = \frac{1}{2} \text{STr}\,\left[\left( \Gamma_k^{(2)} +R_k \right)^{-1}\p_t R_k\right],
\label{Wetterich equation}
}
which is known as the Wetterich equation, see also \cite{Morris:1993qb}, and  \cite{Berges:2000ew, Aoki:2000wm,Polonyi:2001se,
Pawlowski:2005xe, Gies:2006wv, Delamotte:2007pf, Rosten:2010vm, Braun:2011pp} for reviews.
Here, $t:=\ln\fn{k/\Lambda}$ with a reference scale $\Lambda$ and $\Gamma_k^{(2)}+R_k$ is the full regularized inverse propagator with a cutoff profile function $R_k\fn{p}$.
In this paper, we employ the Litim-type cutoff function~\cite{Litim:2001up}:
\al{
r_k\fn{p}=\paren{k^2-p^2}\theta\fn{k^2-p^2},
}
which is multiplied by the wave-function renormalization and an appropriate tensor structure for each field such that in the full regularized propagator $(\Gamma_k^{(2)}+R_k)^{-1}$ the momenta are replaced by $k^2$.\\
In this way, the path integral given as a functional integral is written as a functional differential equation with a boundary condition. The boundary condition is provided by specifying the effective action at a cutoff scale $k$. Exploring whether a model is asymptotically safe can be understood as the search for a consistent boundary condition for which the limit $k\rightarrow \infty$ can be taken and for which the flow to the infrared (IR) depends on a finite number of free parameters. 
The Wetterich equation reproduces one-loop perturbation theory straightforwardly; the extraction of higher-loop orders in discussed, e.g., in \cite{Papenbrock:1994kf,Codello:2013bra}. Within its one-loop structure, it encodes effects beyond perturbation theory, as it depends on the full, field- and momentum dependent propagator. Accordingly, it is particularly well-suited to study interacting fixed points which require nontrivial resummation techniques to be accessible with perturbation theory.

Although the Wetterich equation is exact, in practice its solution requires making approximations.
 We restrict the theory space to a subspace with a finite number of effective operators as an approximation. Guidance to set up reliable truncations is provided by the canonical dimension of couplings, which determines whether a coupling is relevant in perturbation theory, and appears to remain a useful guiding principle at an asymptotically safe fixed point in gravity \cite{Falls:2013bv}, as well as in gravity-matter systems, see, e.g., \cite{Narain:2009fy,Eichhorn:2016esv,Eichhorn:2016vvy,Eichhorn:2017sok}.

For a given effective action, one can obtain the beta functions using Eq.~\eqref{Wetterich equation},
\al{
\p_t {\tilde g}_i =\tilde \beta_i\fn{\{\tilde g\}},
}
where ${\tilde g}_i=g_ik^{-d_i}$ are dimensionless couplings and $\{\tilde g\}=\{{\tilde g}_1,...,{\tilde g}_n\}$ denotes a set of them;
and $d_i$ are the canonical dimensions of the dimensionful couplings $g_i$.
The beta functions can be written as
\al{
\tilde \beta_i\fn{\{\tilde g\}}=-d_i\, {\tilde g}_i + f_i\fn{\{{\tilde g}\}},
\label{explicit beta function}
}
where the first term is the canonical scaling term and the second term arises from loop effects.
At a fixed point ${\tilde g}_i^*$ all beta functions vanish, i.e., $\tilde \beta_i\fn{\{{\tilde g}^*\}}=0$.
In the vicinity of a fixed point, we can classify the directions of the RG flow into UV or IR repulsive or attractive.
To this end, let us expand the beta functions around the fixed point:
\al{
\p_t {\tilde g}_i 
&= \tilde \beta_i\fn{\{{\tilde g}^*\}} + \frac{\p \tilde \beta_i}{\p {\tilde g}_j}\bigg|_{{\tilde g}={\tilde g}^*}({\tilde g}_j-{\tilde g}_j^*)+\cdots\nn
&\simeq \mathcal{M}_{ij}
({\tilde g}_j-{\tilde g}_j^*).\label{eq:linflow}
}
The first term on the right-hand side vanishes by definition of the fixed point.
The solution to the beta functions is given by
\al{
{\tilde g}_i = {\tilde g}_i^* +\sum_j C_j V_i^j \left(\frac{k}{\Lambda_0} \right)^{-\theta_j},\label{eq:linrun}
}
where $V_j$ is an eigenvector of the stability matrix $\mathcal{M}_{ij}$ and $C_j$ are arbitrary constants of integration.
The values $-\theta_j$ are eigenvalues of $\mathcal{M}_{ij}$ and are called critical exponents.
For positive critical exponents, the RG flow toward the IR, i.e., to lower $k$, goes away from the fixed point $g_i^*$.
The corresponding operators are relevant and the IR value of the corresponding superposition of couplings parameterizes the deviation from scale-invariance. Accordingly, a free parameter, corresponding to the choice of $C_j$, fixed by comparison to experiment, is associated to each relevant coupling.
In contrast, the RG flow with negative critical exponents is pulled toward the fixed-point value toward the IR and the corresponding couplings are irrelevant. Beyond the linear approximation in Eq.~\eqref{eq:linflow}, the value of irrelevant couplings changes as a function of scale, but there is no free parameter associated to it. The absence of a free parameter for each irrelevant operator can also be understood by thinking about the flow toward the UV: The flow can only end up at the fixed point in the UV if it stays exactly within the critical hypersurface, which is spanned by the UV attractive directions. According to Eq.~\eqref{eq:linrun}, these are the relevant couplings. Therefore, there cannot be a free parameter associated to an irrelevant coupling. In other words, if $C_j\neq 0$ is chosen for an irrelevant direction, this leads to a flow that will deviate from the critical hypersurface toward the UV, and will not result in a UV complete trajectory.

In the asymptotic safety scenario, the UV complete theory is given by the UV critical surface spanned by the relevant operators. If it features a finite number of positive critical exponents then the model is predictive and low energy physics is determined by the values of the relevant couplings at some scale. 
At an interacting fixed point, one can argue that only a finite number of relevant couplings should exist.
Using Eq.~\eqref{explicit beta function} the critical exponents read
\al{
\theta_i \simeq d_i - \frac{\p f_i}{\p {\tilde g}_i}\bigg|_{{\tilde g}={\tilde g}^*},
}
where the off-diagonal parts of $\mathcal{M}_{ij}$ are neglected for simplicity.
The first term corresponds to the canonical dimension of the coupling. 
The second one arises from loop effects and provides a finite shift of the scaling dimension at an interacting fixed point away from the canonical dimension.
Note that perturbation theory corresponds to the dynamics around the Gaussian fixed point $\tilde g^*=0$ at which the critical exponents are given by the canonical dimension of coupling constants, $\theta_i\simeq d_i$.
It is essential for the asymptotic safety scenario that nonperturbative dynamics around a nontrivial fixed point ${\tilde g}^*\neq 0$ generates nontrivial anomalous dimensions.
For asymptotically safe quantum gravity and matter, results within most truncations suggest that quantum corrections are not large, such that only the Newton coupling, the cosmological constant, and a superposition of couplings at second order in the curvature are relevant, but all other gravity couplings remain irrelevant \cite{Codello:2008vh,Benedetti:2009rx,Falls:2013bv,Denz:2016qks,Gies:2016con}. This provides us with 
 the rationale to choose truncations according to the canonical dimension, as we expect that couplings which are irrelevant according to their canonical dimension will not be shifted into relevance.

\subsection{Effective action}
We investigate the following truncated effective action in four dimensional Euclidean spacetime:
\al{
\Gamma_k=\Gamma_k^\text{EH}+\Gamma_k^\text{matter},
\label{general action}
}
where the gravity sector is given by the Einstein-Hilbert truncation, namely,
\al{
\Gamma_k^\text{EH}[g]
	&=	\frac{1}{16\pi G}\int\df^4x \sqrt{g}[-R + 2\Lambda]
		+S_{\rm gf}	+S_{\rm gh},
		\label{effective action for gravity}
}
where $G$ and $\Lambda$ are the Newton constant and the cosmological constant, respectively; $R$ is the Ricci scalar; $S_{\rm gf}$ and $S_{\rm gh}$ are the actions for the gauge fixing and ghosts whose forms are given below.
The matter sector contains two real scalar fields,
\al{
\Gamma_k^\text{matter}&[\phi, \chi]
	=	\int\df^4x \sqrt{g}\Bigg[ V\fn{\phi,\chi}\nn
&	+\frac{Z_{k,\phi}}{2} g^{\mu \nu}\,\p_\mu{\phi}\,\p_{\nu}\phi 
		+\frac{Z_{k,\chi}}{2} g^{\mu \nu}\,\p_\mu{\chi}\,\p_ {\nu}\chi
		\Bigg],
				\label{effective action for matter}
}
where $\phi$ is a field associated with the massive mode of the Higgs boson, while $\chi$ is a single scalar boson. 
In the SM, the Higgs field is a complex SU(2) doublet, but here
we neglect the Goldstone bosons and use a $\mathbb{Z}_2$ symmetric real scalar as a toy model for the Higgs.
For the second scalar, we also impose a $\mathbb{Z}_2$ symmetry, which prohibits interaction terms of uneven powers in the fields that could lead to a potential that is not bounded from below.
Moreover, we neglect additional degrees of freedom of the SM; most importantly the fermions which can provide a direct contribution to the flow of the Higgs potential and impact the fixed-point values for $G, \Lambda$. In our analysis, we will focus on a fixed point that preserves the $\mathbb{Z}_2$ symmetry as well as shift symmetry in both scalar fields, and accordingly features a vanishing potential. 

The scalar potential is expanded into  polynomials of the fields such that it respects the $\mathbb{Z}_2$ symmetries
, that is,
\al{
V\fn{\phi,\chi}&=\frac{m_\phi^2}{2}\phi^2+\frac{\lambda_\phi}{8}\phi^4
+\frac{\lambda_{\phi \chi}}{8}\chi^2\phi^2
+\frac{m_\chi^2}{2}\chi^2+\frac{\lambda_\chi}{8}\chi^4.
}
The higher-order terms are set to zero since their canonical dimensions are negative, and thus we expect that they are irrelevant even though gravitational fluctuations are taken into account. 
Note that the action for the matter sector \eqref{effective action for matter} is symmetric under the exchange of $\phi$ with $\chi$.

In our truncation, the effective action is parametrized by seven couplings, namely, the Newton constant $G$, the cosmological constant $\Lambda$, the scalar field masses $m_{\phi,\chi}$, the quartic couplings $\lambda_{\phi,\chi}$, and the Higgs portal coupling $\lambda_{\phi \chi}$ and the anomalous dimensions $\eta_{\phi,\chi}$ which are related to the wave-function renormalizations $Z_{\phi,\chi}$ via $\eta_{\phi,\chi} = - \partial_t \ln Z_{\phi,\chi}$.
To obtain the beta functions for the system \eqref{general action}, we employ the background field method.
To this end, we perform a linear split of the metric into a background metric and a fluctuation:
\al{
g_{\mu\nu}={\bar g}_{\mu\nu} + \sqrt{32\pi G Z_h}h_{\mu\nu},\label{eq:backgroundsplit}
}
where $Z_h$ is the graviton wave-function renormalization and the associated anomalous dimension is given by $\eta_h = -\partial_t \ln Z_h$.
Note that $h_{\mu\nu}$ is not restricted to be small in amplitude, i.e., Eq.~\eqref{eq:backgroundsplit} is not a perturbative expansion.

The gauge fixing and the ghost action are given by
\al{
S_{\rm gf}	&=	\frac{1}{2\alpha}\int \df^4x\sqrt{\bar g}\,
				{\bar g}^{\mu \nu}\Sigma_\mu\Sigma_{\nu},
					\label{gaugefixedaction} \\
S_{\rm gh}	&=	-\int\df^4x\sqrt{\bar g}\,\bar C_\mu\left[ {\bar g}^{\mu\rho}{\bar \nabla}^2+\frac{1-\beta}{2}{\bar \nabla}^\mu{\bar \nabla}^{\rho}		+{\bar R}^{\mu\rho}\right] C_{\rho}, \label{ghostaction}
}
with
\al{
\Sigma_\mu		
	:= {\bar \nabla}^\nu h_{\nu \mu}-\frac{\beta +1}{4}{\bar \nabla}_\mu h,
}
where $\bar{\nabla}_\mu$ is the covariant derivative with respect to the background metric; $h:={\bar g}_{\mu\nu}h^{\mu\nu}$ is the trace mode of $h_{\mu\nu}$; $C$ and $\bar C$ are the ghost and antighost fields, respectively; and $\alpha$ and $\beta$ are gauge parameters.

\subsection{Structure of beta functions}
To search for a scale-invariant fixed-point regime, we make a transition to dimensionless couplings, defining
\al{
{\tilde G}&= Gk^2,&
{\tilde \Lambda}&=\Lambda k^{-2},&
}
whose beta functions are
\al{
\p_t {\tilde G}&
=2{\tilde G} +f_{G},&
\p_t {\tilde \Lambda}&
=-2{\tilde \Lambda} +f_{\Lambda}.&
}
In the matter sector, we define dimensionless renormalized fields,
\al{
{\tilde \phi}&= \frac{Z^{1/2}_\phi \phi}{k},&
{\tilde \chi}&= \frac{Z^{1/2}_\chi \chi}{k}.&
}
For the effective potential, we have $k^4\tilde V({\tilde \chi,\tilde \phi})=V\fn{\chi,\phi}$, which implies
\al{
V\fn{\chi,\phi}
&=\frac{m_\phi^2}{2}\phi^2+\frac{\lambda_\phi}{8}\phi^4
+\frac{\lambda_{\phi \chi}}{8}\chi^2\phi^2
+\frac{m_\chi^2}{2}\chi^2+\frac{\lambda_\chi}{8}\chi^4\nn
&=k^4\Bigl[\frac{{\tilde m}_\phi^2}{2}{\tilde \phi}^2+\frac{{\tilde m}_\chi^2}{2}{\tilde \chi}^2\nn
&\qquad+\frac{{\tilde \lambda}_\phi}{8}{\tilde \phi}^4
+\frac{{\tilde \lambda}_{\phi \chi}}{8}{\tilde \chi}^2{\tilde \phi}^2
+\frac{{\tilde \lambda}_\chi}{8}{\tilde \chi}^4\Bigr]\nn
&=k^4{\tilde V}(\tilde \chi, \tilde \phi),
}
where the dimensionless renormalized couplings are defined by
\al{
{\tilde m}_\phi^2&= \frac{m_\phi^2}{Z_\phi k^2},&
{\tilde m}_\chi^2&=\frac{m_\chi^2}{Z_\chi k^2},&
{\tilde \lambda}_{\phi \chi}&=\frac{\lambda_{\phi \chi}}{Z_\phi Z_\chi},&\nn
{\tilde \lambda}_\phi&= \frac{\lambda_\phi}{Z_\phi^2},&
{\tilde \lambda}_\chi&= \frac{\lambda_\chi}{Z_\chi^2}.&
}
Then, the beta functions for dimensionless coupling constants are 
\al{
\label{beta function of mphi}
{\tilde \beta}_{m_{\phi}^2}
&=(-2+\eta_{\phi}){\tilde m}_{\phi}^2+{f}_{m_{\phi}^2} ,\\
{\tilde \beta}_{m_{\chi}^2}
&=(-2+\eta_{\chi}){\tilde m}_{\chi}^2 + {f}_{m_{\chi}^2},\\
{\tilde \beta}_{\lambda_{\phi \chi}}
&=(\eta_\phi + \eta_\chi){\tilde \lambda}_{\phi \chi} + {f}_{\lambda_{\phi \chi}}\\
{\tilde \beta}_{\lambda_{\phi}}
&=2 \eta_{\phi} {\tilde \lambda}_{\phi}+ {f}_{\lambda_{\phi}},\\
\label{beta function of lambdachi}
{\tilde \beta}_{\lambda_{\chi}}
&=2 \eta_{\chi} {\tilde \lambda}_{\chi}+ {f}_{\lambda_{\chi}}.
}
$f_{g}$ are the one-loop corrections computed with the Wetterich equation \eqref{Wetterich equation}.
The RG equations are given by $ \p _t\tilde g_i=\tilde \beta_{g_i}$.
The explicit forms for the dimensionless beta functions for the matter coupling constants are shown in Appendix~\ref{explicit forms of beta functions}. 
For the scalar subsector, these agree with the perturbative one-loop result, once the threshold-corrections from the FRG are set to zero. These are responsible for an automatic decoupling of massive modes, once the RG scale falls below the mass of those modes.

\section{Results}
The beta functions in our truncation feature a fixed point at finite gravitational couplings with an exactly vanishing scalar potential, in accordance with the symmetry considerations in \cite{Eichhorn:2017eht}.  These guarantee that the hypersurface with unbroken shift symmetry in the scalars is a fixed hypersurface under the RG flow. 
For the scalar subsector, this result follows from the well-known fact that global symmetries of the action which are preserved by the regularization remain symmetries at the quantum level. Within asymptotically safe gravity, the same appears to hold when quantum fluctuations of gravity are included. Interestingly, a similar result appears to hold in the effective field theory regime for quantum gravity, where gravity-corrections to the quartic coupling vanish unless a finite scalar mass is present, see \cite{Rodigast:2009zj,Mackay:2009cf}.
Momentum-dependent gravity-induced scalar couplings  that respect shift-symmetry and are necessarily finite at a joint fixed point of the system \cite{Eichhorn:2012va,Eichhorn:2017sok} are not included in our truncation, which therefore features a Gaussian matter fixed point in analogy to the system with one scalar, \cite{Narain:2009fy}.  In addition to the shift-symmetric fixed point, a fixed point with explicitly broken shift-symmetry could of course exist, but in our truncation no such fixed point with a potential that is stable in our simple polynomial approximation is discovered.
As an example for the numerical results, we choose the gauge parameters as $\alpha\to 0$ and $\beta=1$.
In this case, the anomalous dimensions of the scalar fields vanish $\eta_\phi=\eta_\chi=0$ for the Gaussian-matter fixed point in symmetric phase.

First, we look for the fixed point at which all beta functions in the system vanish; $\tilde \beta_g\fn{\{g^*\}}=0$.
We find the Gaussian-matter fixed point, namely, only the Newton coupling and the cosmological constant have a nonvanishing fixed point value,
\al{
\tilde G^*&=1.182,&
\tilde \Lambda^*&= 0.161,&
\label{eq:NGFPvalue}
}
while the matter couplings vanish, where we set $\eta_h=0$.
At this fixed point, the critical exponents take the following numerical values:
\al{
&\theta_{1,2}=2.5083 \pm 1.6384 i,&
&\theta_{3,4}=-0.45478,&\nn
&\theta_{5,6,7}=-2.4548.&
}
Here, $\theta_{1,2}$ are the critical exponents associated to the two relevant directions located in the Einstein-Hilbert subspace. The effective scaling of $G$ and $\Lambda$ sensibly deviates from the canonical scaling induced by a Gaussian fixed point. Therefore, the non-Gaussian fixed point \eqref{eq:NGFPvalue} has nonperturbative origins; $\theta_{3,4}$ correspond to the scalar masses; $\theta_{5,6,7}$ correspond to the quartic couplings and the Higgs portal coupling.
At this fixed point, all terms in the scalar potential are irrelevant, i.e., the nonperturbative quantum-gravity effects are strong enough to render the mass parameters irrelevant, even though their canonical dimension is 2. Accordingly, the low-energy form of the potential is fully determined in terms of the IR-values of the gravitational couplings. In particular, within our truncation the potential stays exactly flat at all scales.

\begin{figure}[!t]
\includegraphics[width=0.9\linewidth]{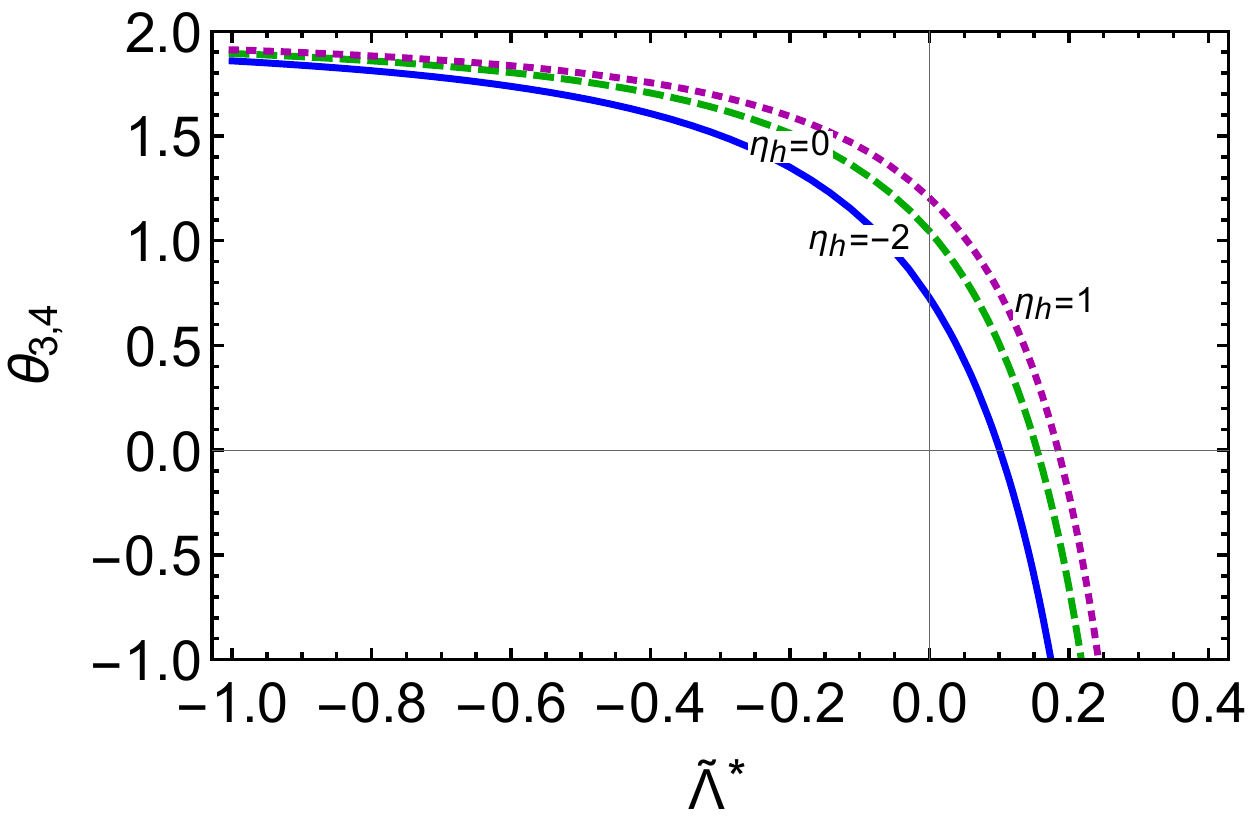}\\ \includegraphics[width=0.9\linewidth]{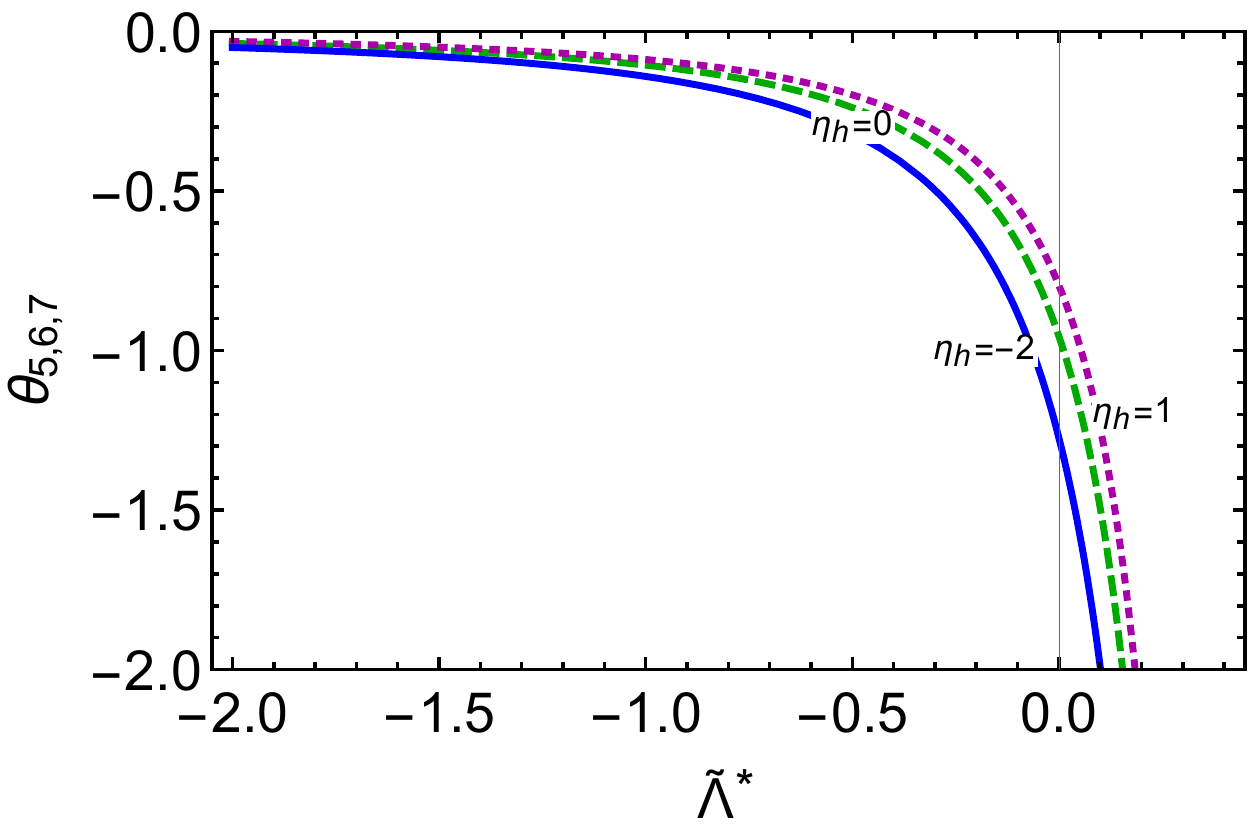}
\caption{\label{fig:critexp} Critical exponents for the mass  and the quartic couplings at $\tilde{G}^{\ast}=1$ as a function of $\tilde{\Lambda}^{\ast}$ for different choices of anomalous dimensions. Here we set the gauge parameters $\beta=1, \alpha=0$. For the choice $\beta=0$, the dependence on $\tilde{\Lambda}^{\ast}$ changes only very slightly.}
\end{figure}

Let us now broaden our view beyond the current truncation and treat $\tilde{G}^{\ast}, \tilde{\Lambda}^{\ast}$ as free parameters. This accounts for extensions of the truncation in the gravity sector. Moreover, we currently employ the single-metric approximation to evaluate $\tilde{G}^{\ast}, \tilde{\Lambda}^{\ast}$, whereas actually, only fluctuation-field couplings should appear on the right-hand-side of the Wetterich equation. Varying $\tilde{G}^{\ast}, \tilde{\Lambda}^{\ast}$ away from their fixed-point values in our approximation allows us to explore whether the system might behave in a qualitatively different way in extended truncations. Further, $\Lambda$ should be viewed as a simple approximation of nontrivial threshold behavior in the full gravity propagator, i.e., varying $\Lambda$ mimics the effect of higher-order terms in the propagator, see, e.g., \cite{Hamada:2017rvn,Eichhorn:2017eht}.
Last but not least, the addition of further matter degrees of freedom, e.g., those of the SM, also results in a change of the fixed-point values in the gravity sector.

\begin{figure}[!t]
\includegraphics[width=0.9\linewidth]{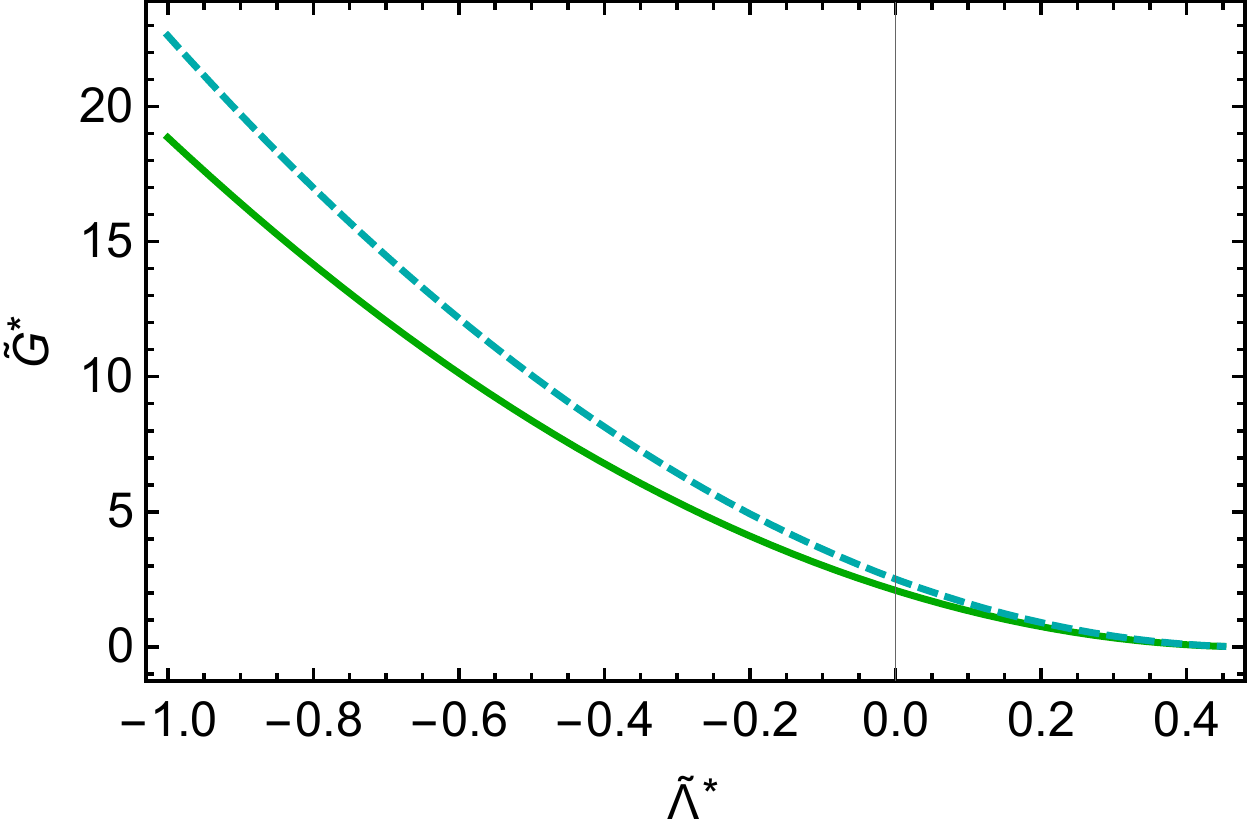}
\caption{\label{fig:critexpGlambda} Above the thick green (cyan dashed) line, $\theta_{3,4}<0$ holds for $\eta_h=0$ ($\eta_h=1$) with $\alpha=0$, $\beta=1$.}
\end{figure}

We observe that the sign of the critical exponent for the quartic couplings is stable under variations of $\tilde{\Lambda}^{\ast}$ as can be seen in Fig.~\ref{fig:critexp}: For $\tilde{\Lambda}^{\ast}\rightarrow -\infty$, gravity fluctuations are suppressed, and $\theta_{5,6,7}$ approach zero from below. The limit of ``strong" gravity, which is reached when $\tilde{\Lambda}^{\ast}$ approaches the pole in the propagator features an increasingly negative $\theta_{5,6,7}$. On the other hand, the situation differs significantly for $\theta_{3,4}$, which is positive and tends to $\theta_{3,4}\rightarrow 2$ for $\tilde{\Lambda}^{\ast}\rightarrow \infty$, as it should. However, once $\tilde{\Lambda}^{\ast}$ starts to approach the pole in the propagator, the critical exponent switches sign.  By increasing $\tilde{G}$, which strengthens gravitational fluctuations, the onset of irrelevance for the mass parameters is shifted to negative $\tilde{\Lambda}$, cf.~Fig.~\ref{fig:critexpGlambda}.

In our approximation, there are two physically distinct regions of the gravitational parameter space: For $\tilde{\Lambda}^{\ast}$ below a critical value, the quartic couplings are irrelevant at the free fixed point, while the masses are relevant. Therefore, the masses remain free parameters, and their IR values can be chosen arbitrarily. On the other hand, the second part of parameter space features irrelevant masses and quartic couplings. Thus, the potential in the scalar sector is completely flat in this case. We observe that the regime in which this holds shrinks as $\eta_h$ is taken to larger values. In this context, we remark that $\eta_h=-2$ holds in the single-metric approximation, while $\eta_h \geq 0$ typically appears as a result from fluctuation calculations. 
Within the background approximation, fixed-point values for gravity under the impact of minimally coupled matter degrees of freedom fall into the regime $\tilde{\Lambda}^{\ast}<0$, where the mass parameters remain relevant \cite{Dona:2013qba}.

 As further work is necessary to establish whether the fixed point lies at positive or negative $\theta_{3,4}$,
we consider two scenarios (i): $\theta_{3,4}<0$ (scenario A) and (ii) $\theta_{3,4}>0$ (scenario B) in the next section.

\section{Potential phenomenological implications}
We now discuss  potential phenomenological consequences of our results in a setting where $\chi$ is interpreted as a dark matter candidate, and our model is a toy model of the Higgs portal to scalar dark matter. Since the Higgs portal coupling becomes irrelevant, the 
dark matter within the present toy model is decoupled from the Higgs sector  at all scales: In accordance with the discussion in \cite{Eichhorn:2017eht}, shift symmetry protects the full potential for the two scalars; thus the gravity-induced fixed point lies at vanishing potential. As the quartic couplings are irrelevant, deviations from shift symmetry in the flow toward the infrared cannot occur in our toy model. In the full SM, additional sources of symmetry-breaking in the Higgs-Yukawa sector, such as, e.g., a finite fixed-point value for the top Yukawa \cite{Eichhorn:2017ylw} or non-Abelian gauge coupling \cite{Harst:2011zx,Eichhorn:2017lry} could lead to a nonzero quartic coupling for the Higgs in the UV, while the vanishing fixed-point value for the Higgs portal coupling remains unaffected by quantum fluctuations of the SM fields. Moreover, even starting from a vanishing Higgs quartic coupling at the Planck scale, top-quark and gauge boson fluctuations build up a nontrivial Higgs potential in the flow toward the IR.
In fact, the observed Higgs mass is connected to a near-vanishing Higgs quartic coupling at the Planck scale \cite{Bezrukov:2012sa,Shaposhnikov:2009pv}.
On the other hand, the Higgs portal coupling remains protected by shift symmetry of the dark scalar $\chi$, and thus vanishes \emph{at all scales}.
 One might say that the dark matter sector is even darker than typically assumed for scalar dark matter models. A decoupled dark sector and the predicted lack of direct and indirect detection 
 appear to be
in line with experiments, where searches have until now succeeded in placing strong bounds on the allowed parameter space, but have not resulted in a detection \cite{Cline:2013gha,Beniwal:2015sdl,Aprile:2015uzo,Akerib:2016vxi,Aprile:2017iyp,Cui:2017nnn}. 
In this setting, dark matter cannot be a thermal relic  as it completely decouples from the SM at all scales, and non-thermal production processes in the early universe need to be invoked. These rely on a nonvanishing dark matter mass, as present in scenario A.

\subsection{Effective theory with near-fixed-point scaling}
Within asymptotic safety, scenario A is incompatible with $\chi$ being a dark matter candidate, as all trajectories emanating from the UV fixed point have a vanishing mass for $\chi$ at all scales.
For the remainder of this subsection, we will thus broaden our view beyond the asymptotic-safety paradigm. Instead we will consider a setting where our analysis is assumed to hold for a range of scales $\Lambda_{\rm UV}\geq k \geq M_{\rm Planck}$, but new physics 
exists
 beyond $\Lambda_{\rm UV}$. Then, the fixed point that we discover is strongly IR attractive in the two mass-parameters, provided the values of the gravitational couplings remain in the regime pertaining to scenario A. Accordingly, the flow is likely to pass close to the fixed point, starting from a whole range of initial conditions at $\Lambda_{\rm UV}$, so that the mass parameters will be close to zero, but \emph{not} exactly vanishing in the vicinity of $M_{\rm Planck}$.

\subsubsection{Decoupled dark sector}
 The full effective action, evaluated at tree level, contains the strength of all possible interactions. In our case, the Higgs portal coupling to dark matter will be zero in the full effective action. This does not yet preclude the existence of dark-matter-Higgs interactions, as, in accordance with shift-symmetry, gravity generates nonvanishing momentum-dependent interactions \cite{Eichhorn:2012va}, potentially allowing for the production of dark matter \cite{Garny:2015sjg,Tang:2016vch,Garny:2017kha}.
The momentum-dependent interactions are canonically irrelevant and are expected to remain irrelevant at their shifted Gaussian fixed point. 
In \cite{Garny:2015sjg,Tang:2016vch,Garny:2017kha}, based on a calculation using the Einstein gravity action, it is shown that a sufficient amount of dark matter can be produced even if there are no interactions between the dark matter and SM particles except for gravity which mediates a momentum-dependent interaction.
Whether their calculation is modified in our case is an intriguing question that we leave open in this study.

An alternative possibility is that the dark matter abundance is explained by the coherent oscillation of the $\chi$ field through the misalignment mechanism, as in the axion dark matter scenario (see Ref.~\cite{Marsh:2015xka} for a recent review). This mechanism relies on the dark matter mass being nonzero.
The misalignment mechanism, as discussed in \cite{Jaeckel2012},  starts from
a spatially homogeneous but time-dependent initial field value $\chi_i\gg 0$ after inflation. 
Assuming a flat Robertson Walker universe, $\chi$ obeys
\begin{equation}\label{eq:KleinGordonFRW}
\ddot{\chi}+3H\dot{\chi}+m_{\chi}^2\chi=0,
\end{equation}
where $H$ denotes the Hubble scale.
In our scenario the irrelevance of interactions between the dark matter field and the thermal bath of SM particles implies that the mass does not receive any thermal corrections and can therefore be regarded as temperature- and hence time-independent.
 Initially, for
$H\gg m_{\chi}$, the solution to Eq.\,\eqref{eq:KleinGordonFRW} is given by an overdamped harmonic oscillator 
\begin{equation}
\chi\fn{t}=\chi_1+\chi_2\,e^{-3Ht},
\end{equation}
such that $\chi$ remains exponentially frozen
to the initial field value
value $\chi_1$. 
At $3 H\fn{t_\chi}=m_\chi$ the system undergoes a crossover and $\chi\fn{t}$ rolls down the quadratic potential well to reach a stable equilibrium point at which the field begins a rapid oscillation. The crossover occurs at a temperature
\al{
T_\chi&=\sqrt{M_P m_\chi \sqrt{90\over\pi^2 g_*(T_\chi)}}\nn
&\sim 10^3\eV \sqrt{m_\chi\over10^{-22}\eV}\paren{1\over g_*}^{1/4},
\label{crossover temperature}
}
where $g_*$ denotes the effective degrees of freedom of the energy density, and we have used that $H=\sqrt{\pi^2g_*\over90}{T^2\over M_P}$ in the radiation dominated era. 
At the time when $H\lesssim m_\chi$, the mass dominates the time-evolution of the field according to
\begin{equation}\label{eq:DarkMatterOscillator}
\chi\fn{t}
\approx \chi_1\left(\frac{a_\chi}{a\fn{t}}\right)^{3/2}\cos\bigg(m_{\chi} \cdot\big(t\fn{T}-t_\chi\fn{T_\chi}\big)\bigg), 
\end{equation}
where $a\fn{t}$ denotes the scale factor at time $t$ and $a_\chi$ is the scale factor at the crossover time $t_\chi$.
The energy density of $\chi$ is then given by
\al{
\rho_\chi\fn{T_\chi}\sim \frac{1}{2}m_\chi^2 \mathcal{A}^2\fn{T_\chi},
\label{dark matter amplitude}
}
where $\mathcal{A}\fn{T}=\chi_1\left(a_\chi/a\fn{T}\right)^{3/2}$ is the amplitude of the oscillation $\chi(T)$ at temperature $T$, which simplifies at the crossover temperature $T_\chi$ to $\mathcal{A}\fn{T_\chi}=\chi_1$.
To estimate for which dark matter mass this process generates the full dark matter abundance  inferred from observations, consider the energy density in a comoving volume, $\rho a^3$ and the entropy in a comoving volume, $sa^3$. Since the mass does not depend on time, Eq.~\eqref{eq:DarkMatterOscillator} and $\rho_\chi(T)\sim1/2\, m_\phi^2 \mathcal{A}^2\fn{T}$ imply that the energy density in a comoving volume is conserved. Conservation of comoving entropy follows from the assumption that the universe expands adiabatically. The two quantities being conserved, it follows that their ratio is also conserved such that the following relation holds:
\al{
{\rho_\chi\fn{T_\chi}\over s\fn{T_\chi}}={\rho_\chi\fn{T_0}\over s\fn{T_0}},
\label{conserved relation}
}
where $T_0$ is the present temperature of the universe and the scale factors cancelled out.
Roughly speaking, the relation \eqref{conserved relation} says that the ratio of the number of the dark matter particles to that of photons is conserved.
Using \eqref{dark matter amplitude} and $s\fn{T_\chi}=2\pi g_{*s}\fn{T_\chi}T_\chi^3/45$, 
we can rewrite the relation \eqref{conserved relation} as
\al{
{\frac{1}{2}m_\chi^2 \chi_1^2\over T_\chi^3 }\simeq {\rho_\chi\fn{T_0}\over s\fn{T_0}},
}
where we assumed that the coefficient of the entropy density is of order one.
Furthermore, we replace the temperature $T_\chi$ by Eq.\,\eqref{crossover temperature}, and on the right-hand side we insert the observed quantities~\cite{Patrignani:2016xqp},  
\al{
 {\rho_\chi\fn{T_0}\over s\fn{T_0}}
&\simeq \frac{1.2497\times 10^{-6}\,\text{GeV}/\text{cm}^{3}}{2891.2\,\text{cm}^{-3}}\nn
&\simeq 4.32\times 10^{-10}\,\text{GeV}.
}
Then, the mass $m_\chi$ is  determined in terms of the initial field value $\chi_i$
\al{\label{Eq: DM mass2}
m_\chi
\sim 10^{-20}\eV \paren{10^{17}\GeV \over \chi_1}^{4},
}
where $g_*(T_\chi)\sim1$ is taken.
Therefore, if the initial amplitude is close to the Planck scale, $m_\chi$ becomes extremely small.
This class of dark matter is called fuzzy dark matter, see
\cite{Hui:2016ltb} for observational constraints.
The observation of the Lyman-$\alpha$ forest puts a lower bound on the mass:
$m_\chi\gtrsim 10\text{--}20 \times10^{-22}\eV$~\cite{Viel:2013apy,Hui:2016ltb}, which implies that the initial amplitude $\chi_1$ should be smaller than $10^{17}\GeV$.
As discussed in \cite{Jaeckel2012},
$\chi$
behaves like a cold dark matter candidate,  as its
equation of state  is that of nonrelativistic matter. Denoting the average over a full oscillation by $\langle \,\rangle$,
\begin{equation}
 w=\frac{\langle p_{\chi}\rangle}{\langle \rho_{\chi} \rangle},
\end{equation}
 with
\begin{equation}
\langle\rho_{\chi}\rangle=\frac{1}{2} m_{\chi}^2\mathcal{A}^2 +\mathcal{O}(\dot{\mathcal{A}}),\quad \langle p_{\chi} \rangle= \frac{1}{2}\dot{\mathcal{A}}^2(t).
\end{equation}
Terms involving a derivative of $\mathcal{A}(t)$ are proportional to $H \mathcal{A}(t)$. Therefore, in the late universe, when $H\ll m_\chi$, these terms are negligible compared to $m_\chi \mathcal{A}(t)$ and the equation of state modifies to
\begin{equation}
w\approx 0,
\end{equation}
which is the equation of state for nonrelativistic matter. 

\subsubsection{The resurgence mechanism}
Depending on the initial conditions, the dark-matter mass required to produce the observed dark-matter abundance via the misalignment mechanism can be rather small compared to the Planck scale. Here, we will highlight that the negative critical exponent of the mass parameter in the quantum-gravity regime can accommodate such a hierarchy in a ``natural" way.
To that end, let us
review some well-known aspects of the quadratic divergences which are associated to the mass parameters in a perturbative setting.
The loop-corrections to the scalar mass involve quadratic divergences which  
depend on regularization schemes, and are not present in dimensional regularization.  In fact, the presence or absence of quadratic divergences depends on a choice of the coordinates of theory space since a choice of the regularization scheme corresponds to specifying a set of coordinates in theory space. Physics must of course be independent of regularization schemes (choices of coordinates in theory space), and simply encoded in different ways in different schemes.
The viewpoint taken in~\cite{Wetterich:2016uxm,Wetterich:1983bi,Wetterich:1990an,
Wetterich:2011aa,Aoki:2012xs} is that 
the position of the phase boundary between the symmetry broken and symmetric phases in the theory space is encoded in a scheme-dependent value of the dimensionless mass parameter ${\bar \mu}^2$.
The deviation from the phase boundary, could be a physical quantity since it does not depend on a choice of coordinates on theory space (on the regularization scheme).  In our parameterization of the scalar potential, the deviation from the phase boundary corresponds to the mass of the scalar, i.e., $\bar{\mu}^2=0$.
It should be noted here that the physics on the phase boundary corresponds to the massless theory. This fact was pointed out by Wetterich in~\cite{Wetterich:1983bi} and can lead to a scale invariant theory. We will discuss the possibility of the scale invariance within the
present extension in Sec.~\ref{Asymptotic safety and Classical scale invariance}.
The RG flow of the deviation from the phase transition $\tilde m^2=m^2/k^2$ is given by
\al{
{\tilde m}^2\fn{k}={\tilde m}_0^2\left( \frac{k}{M}\right)^{-\theta_m},
\label{RG flow of scalar mass}
}
where ${\tilde m}_0^2 = \tilde{m}^2(k=M)$.
This RG equation is obtained from \eqref{eq:linrun} with $m_*^2=0$; $V_i^j=\delta^j_i$; $C_j=\tilde m_0^2$; and $\Lambda_0=M$.
Since the scalar mass in the SM is relevant and its critical exponent is approximately $\theta_m\approx 2$, the scalar mass at the Planck scale $M=M_\text{P}$ has to be 
much
smaller than one, namely, ${\tilde m}_0=m\fn{M_P}/M_P\ll 1$.
This is the gauge hierarchy problem~\cite{Gildener:1976ai,Weinberg:1978ym}. Phrased in physical terms, the question is why the SM lies so close to the phase boundary at microscopic scales.
The resurgence mechanism was suggested in Ref.~\cite{Wetterich:2016uxm} as a solution for this problem within asymptotically safe gravity.
It relies on the negative critical exponent of the scalar mass ($\theta_m<0$), 
generated by quantum fluctuations of gravity. 

The resurgence mechanism links the physical mass of the dark matter scalar to a ``natural" UV cutoff scale.
Loosely speaking, the scaling dimension of $\tilde m_\chi^2$ is $\theta_4\simeq 2$ below the Planck scale, and is $\theta _4\simeq -0.45$ above the Planck scale at least in the vicinity of the fixed-point values for $\tilde{G}, \tilde{\Lambda}$. 
Let us model the critical exponent as 
\al{
\theta_4\simeq 2\theta(\tau-1)-0.45\theta(1-\tau),
\label{approximated form of critical exponent}
}
where $\tau=\frac{1}{8\pi\, G\,k^2}$ is the running reduced Planck scale and $\theta\fn{x}$ is the step function.
The regimes $\tau\gg 1$ and $\tau\ll 1$ correspond to scales below and above the Planck scale, respectively.
Then, the mass squared of $\chi$ at the scale $\Lambda>M_P$ is given by
\al{
{\tilde m}_\chi^2\fn{\Lambda}
={\tilde m}_\chi^2\fn{k_\chi} \paren{k_\chi \over M_P}^2 \paren{M_P\over \Lambda}^{-0.45},\label{eq:mLambda}
}
where $k_\chi$ is the scale at which $m_\chi$ becomes  the physical mass in
\eqref{Eq: DM mass2} and ${\tilde m}_\chi^2\fn{k_\chi}=m_\chi^2/k_\chi^2$ is the dimensionless mass-squared at the scale $k_\chi$. 
Requiring that $\tilde m_\chi^2\fn{\Lambda}\simeq 1$ and ${\tilde m}_\chi^2\fn{k_\chi} \simeq 1$, we obtain the ``natural" cutoff scale,
\al{
\Lambda_\text{natural}\sim \paren{M_P\over m_\chi}^{40/9}M_P.
}
Let us estimate the ``natural" cutoff scale  (i.e., the UV scale at which $\tilde{m}_{\chi}(\Lambda)=1$),  starting from which
the small mass of 
\eqref{Eq: DM mass2} is  obtained in the IR. 
The cutoff scale is
\al{
{\Lambda_\text{natural}\over M_P}\sim 
			 10^{249}\displaystyle \paren{\chi_1\over M_P}^{160/9}		
				\label{natural scale}}
for Eq.~\eqref{Eq: DM mass2}.
The scale \eqref{natural scale} could be regarded as a scale of new physics if we demand that new physics should satisfy a ``naturalness" criterion in the sense of providing dimensionless couplings of order one at the cutoff scale.
Note also that since the dark sector is decoupled from the Higgs sector, there is no relation between the dark matter mass and the electroweak scale.
The natural scale for the Higgs mass above the Planck scale is estimated by the resurgence mechanism with $\theta_3\simeq -0.45$ and is given as $\Lambda_\text{natural EW}\sim 10^{94}\,\text{GeV}$.  
To obtain the observed dark-matter abundance through the misalignment mechanism, while imposing the above ``naturalness" criterion on both the dark matter and the Higgs sector determines the initial field value to be $\chi_1 \simeq 10^{-10}M_P$. As both critical exponents are equal in our toy model, a significant difference between the Higgs mass and the dark matter mass cannot be ``naturally" accommodated.

\subsection{Asymptotic safety}

\subsubsection{Classical scale invariance}\label{Asymptotic safety and Classical scale invariance}
In asymptotic safety, there is no ``scale of new physics," and thus the cutoff scale in Eq.~\eqref{eq:mLambda} is taken to infinity. In scenario A, this provides a completely flat potential at the Planck scale, 
as one can see
from Eq.~\eqref{RG flow of scalar mass}:
In order for the scalar mass with the negative critical exponent to be UV safe, we have to set $\tilde m_0^2=0$, which implies $\tilde m^2\fn{k}=0$ for all values of $k$.
Hence, quantum gravity fluctuations generate a completely flat scalar potential at the Planck scale.\footnote{While we do not explicitly include higher-order terms in the potential here, symmetry considerations imply that their fixed-point values vanish as well. As the direct quantum-gravity contribution to all terms in the scalar potential is the same, the canonically irrelevant higher-order terms are irrelevant at the UV fixed point. Accordingly, the full scalar potential is exactly flat at the Planck scale.}
This setting with so-called classical scale invariance in the scalar sector has been widely explored in the literature~\cite{Hashimoto:2013hta,Hashimoto:2014ela}. ``Classical" here pertains to the microscopic action at the Planck scale, which one might take as the starting point to define a matter model without quantum gravity.
In the present scenario, 
scale invariance---i.e., the absence of dimensionful couplings---in the scalar sector at the Planck scale is an automatic consequence of the dynamics of asymptotically safe gravity.

Models with ``classical scale invariance" have been explored as they might provide a starting point for a dynamical generation of the electroweak scale. 
``Scalegenesis" for the  electroweak (and the dark matter) scale, i.e.,  
the
dynamical generation of these scales, could occur, e.g., by 
dimensional transmutation in the Coleman-Weinberg mechanism~\cite{Meissner:2006zh,Foot:2007iy} or 
through
strong dynamics similar to quantum chromodynamics~\cite{Hur:2011sv,Kubo:2014ova,Kubo:2015cna}.
 Within the SM, the Coleman-Weinberg mechanism is not sufficient to generate the electroweak scale, and additional bosonic fluctuations are required, such as, e.g., a dark matter scalar.
The scenario that a single scalar field could be a dark matter candidate within a classically scale invariant extension of the SM is discussed in \cite{Foot:2010av,Guo:2014bha,Khoze:2014xha,Farzinnia:2014xia,Endo:2015nba}. 
In this case, however, the quartic and Higgs portal couplings have to be relevant which appears to be in tension with an asymptotically safe UV completion within our toy model and truncation thereof.
 In our setting, the full potential is flat at the Planck scale, a scenario known as ``flatland" \cite{Hashimoto:2013hta,Hashimoto:2014ela}.
The flatland scenario has been discussed in \cite{Hashimoto:2013hta,Hashimoto:2014ela}, where a $U(1)_{B-L}$ gauge field (called $Z'$ boson) and a Majorana-type Yukawa interaction between a right-handed neutrino and a singlet complex scalar field are introduced.
However, the singlet complex scalar field cannot be a dark matter candidate since it has a nonvanishing expectation value $\langle \chi \rangle \neq 0$ and then becomes unstable due to its decay into the lighter SM particles. It is an intriguing question that we leave open here whether this model can be rendered asymptotically safe by coupling it to quantum gravity.

\subsubsection{Asymptotic safety in scenario B}
We now consider the case where the critical exponents of the scalar masses become positive.  In this setting, an asymptotically safe model of dark matter with the relic abundance generated from the misalignment mechanism might be viable, as
the mass scale of the dark matter scalar is not determined from the fixed-point-dynamics. 
Just as in the case of any (marginally) relevant coupling, all low-energy values within the basin of attraction of the fixed point are compatible with the requirement that the theory becomes asymptotically safe in the UV. On the other hand, the quartic couplings remain irrelevant, and thus only one distinct low-energy value for each of those quantities, given as a function of the relevant couplings, i.e., $G, \Lambda, m_{\phi}, m_{\chi}$, is compatible with an ultraviolet complete model.\\
Conventionally, a relevant coupling is associated with a fine-tuning problem, whereas a marginally relevant one is not. Note that for both cases there is no way to determine the IR value, and both are sensitive to physics at microscopic scales, in the sense that a change of the value of the coupling at a UV scale by some amount leads to a difference in the IR value. The only distinction lies in the power-law sensitivity of the mass to the cutoff scale in comparison to the logarithmic dependence of a marginally relevant coupling. Within the conventional view on this question, the reduction of the critical exponent by quantum-gravity effects, cf.~Fig.~\ref{fig:critexp} could be viewed as a significant improvement of the situation.
It should be stressed that in any case the fine-tuning ``problem" does not make the theory inconsistent.

Here, we highlight that scenario B is one which appears to make an asymptotically safe UV completion of the Higgs portal to dark matter observationally viable, as finite IR values for the masses are compatible with an asymptotically safe fixed point. Within our toy model and truncation therefore, predictions arising from this fixed point include a vanishing quartic dark-matter coupling, i.e., dark matter is not self-interacting  through momentum-independent interaction channels.
Further, the Higgs mass becomes a prediction, once the electroweak scale is fixed, as the value of the Higgs quartic coupling is predicted, see the discussion in \cite{Shaposhnikov:2009pv} and \cite{Eichhorn:2017ylw} for an explicit construction. Finally, the portal coupling is predicted to vanish. Therefore, direct searches for the scalar dark matter particle would be unsuccessful in this setting. The production of the dark matter particle could proceed via the misalignment mechanism, which would be available in this setting as the mass of the dark matter scalar can be freely chosen in the IR.

Phenomenologically, scenario B is therefore the preferred scenario within asymptotic safety, as a vanishing Higgs portal appears to be observationally viable, while the dark matter mass must be finite, and thus cannot become an irrelevant direction at a free, gravity-induced fixed point.

Let us consider the scalar fields analyzed here in a context beyond the Higgs portal to dark matter. In fact, scalar fields also occur in the context of inflation. Our results might tentatively be interpreted as suggesting that if an inflaton is coupled to asymptotic safety, its potential will generically be flat in the UV. Toward the IR, the mass might be relevant, as in scenario B, or the potential might remain flat, as in scenario A. This would appear to make asymptotic safety in regime A incompatible with inflation driven by an additional scalar field. Scenario B would appear to be still compatible with the data on the inflationary parameters determined by the Planck satellite \cite{Ade:2015lrj}. On the other hand, one might conclude that asymptotic safety appears to disfavor inflation driven by an additional scalar field -- the case of Higgs inflation~\cite{Bezrukov:2007ep} might be an exception.
We stress that this interpretation of our results requires additional extensions of the truncation---here we only discover first hints for such a scenario.
Intriguingly, the microscopic gravity dynamics themselves might drive inflation through higher-order curvature terms \cite{Bonanno:2015fga}, similar to the case of Starobinsky inflation~\cite{Starobinsky:1980te}.

\section{Conclusions and Outlook}
In the present work, we study a model involving a so-called Higgs portal interaction between two real scalar fields, mimicking the Higgs field coupled to a real singlet-scalar dark matter field, under the impact of gravitational fluctuations.
We employ FRG methods and truncate the space of couplings to the canonically marginal and relevant ones. Our explicit results confirm that the canonical dimension does in fact provide a good principle to find consistent truncations, as quantum-gravity effects can even shift the canonically relevant couplings into irrelevance. In this truncation, we find a Gaussian-matter fixed point in agreement with general arguments on the fixed-point structure based on global symmetries \cite{Eichhorn:2017eht}. At the fixed point, all couplings, except for the Newton coupling and the cosmological constant, have a vanishing fixed-point value. 
The Higgs mass, the scalar dark matter mass and all quartic couplings are irrelevant, yielding an exactly flat scalar potential at the Planck scale. 
To extend our truncation in the gravity sector, we consider $\tilde{G}^*$ and $\tilde{\Lambda}^*$ as free parameters. Notably, the Higgs portal coupling stays irrelevant for all $\tilde{\Lambda}^*$. If these results persist beyond our truncation and under the inclusion of additional (beyond) SM degrees of freedom, they hint that a simple scalar dark matter candidate does not couple to the SM through a finite momentum-independent Higgs portal coupling.
Further, this suggests that nonthermal production mechanisms, such as the misalignment mechanism or pure gravitational interactions, could be required to produce the observed dark matter abundance.\\
 We further identify two different scenarios, where the mass is relevant or irrelevant, depending on the fixed-point value of $\tilde{\Lambda}^*$. A scenario with an irrelevant mass is incompatible with $\chi$ being a dark-matter candidate in an asymptotically safe setting.  The second scenario that we identify is characterized by two relevant mass parameters.  If it persists under extensions of the truncation, this could render asymptotic safety compatible with an observationally viable dark matter scalar that has a finite IR mass and can be produced nonthermally, e.g., via the misalignment mechanism.
\\
 We also broaden our view beyond the asymptotic-safety scenario and analyze the system with a finite UV cutoff scale which could be interpreted as the scale of new physics. The resurgence mechanism \cite{Wetterich:2016uxm}, then generates a small scalar mass (in units of the Planck scale) ``naturally" if one starts from a particular UV cutoff scale. This is a consequence of quantum gravity fluctuations rendering the scalar mass parameter \emph{irrelevant}---thus, in contrast to canonical scaling, the dimensionless mass \emph{shrinks} if the momentum scale is lowered in the trans-Planckian regime. Below the Planck scale, where quantum fluctuations of gravity decouple, the dimensionless mass starts to grow toward the IR.
Thus a dimensionless mass of order one at the UV cutoff can become compatible with a tiny IR-mass in units of the Planck scale, as the mass is driven toward zero at the Planck scale by quantum fluctuations of gravity.\\
We further discuss ``classical" scale invariance, in the sense of a flat scalar potential at the Planck scale. Models realizing this condition have been explored as starting points for a dynamical generation of the electroweak scale.  
Within asymptotic safety, this condition is automatically satisfied  in a region of the space of microscopic gravitational couplings, as the scalar mass  features an IR attractive fixed point at zero under the impact of quantum fluctuations of gravity.\\
 Let us emphasize that our results have been obtained within a simple truncation of the scalar and the gravitational sector. 
More specifically, we have neglected higher-order momentum-dependent scalar self-interactions \cite{Eichhorn:2012va} and scalar-curvature interactions \cite{Eichhorn:2017sok} as these do not directly impact the flow of the scalar potential.
On the other hand, they have finite fixed-point values and thereby affect the critical exponents through their effect on the anomalous dimension. 
Within the regime of scenario A, the direct gravity contribution to $\theta_{3,4}$ is expected to dominate \cite{Eichhorn:2017eht}, and thus this particular extension of the truncation does presumably not alter our conclusions pertaining to this regime.
Further, we have neglected higher-order terms in the propagator of metric fluctuations. 
Depending on their fixed-point values, these can lead to different properties of the scalar sector at the fixed point, cf.~\cite{Hamada:2017rvn, Eichhorn:2017eht} for corresponding studies including a Yukawa sector.
Finally, adding further (beyond) SM degrees of freedom can alter the fixed-point structure: For instance, a finite fixed-point value for the Yukawa couplings \cite{Eichhorn:2017ylw} and the gauge couplings \cite{Eichhorn:2017lry,Eichhorn:2017muy}, could generate a nonzero fixed-point potential for the Higgs.

\section*{Acknowledgements}
We acknowledge helpful discussions with J.~Jaeckel.
A.~E.~and J.~L.~are supported by an Emmy-Noether grant of the Deutsche Forschungsgemeinschaft (DFG) under Grant No.\,Ei/1037-1. A.~E.~is also supported by an Emmy-Noether visiting fellowship at the Perimeter Institute for Theoretical Physics. This research was supported in part by Perimeter Institute for Theoretical Physics. Research at Perimeter
Institute is supported by the Government of Canada through the Department of Innovation, Science, and
Economic Development, and by the Province of Ontario through the Ministry of Research and Innovation.
The work of Y.\,H. is supported by the Grant-in-Aid for Japan Society for the Promotion of Science Fellows, No.\,16J06151.
The work of M.\,Y. is supported by the DFG Collaborative Research Centre SFB\,1225 (ISOQUANT). 

\begin{appendix}
\section{Explicit forms of beta functions}\label{explicit forms of beta functions}
We list the explicit forms of the beta functions of the matter coupling constants in the Landau gauge $\alpha=0$ with the gauge parameter $\beta$ left unspecified.
The beta functions of the Newton constant and the cosmological constant have been calculated in many papers; see e.g.,~\cite{Codello:2008vh}.
To show the explicit forms, we define the threshold function: 
\begin{widetext}
\al{
{\mathcal I}\fn{n_{p},n_{h},n_\phi,n_\chi}
= \paren{1-2 \cosmo}^{-n_{p}}\paren{9-6\beta-12\cosmo+\beta^2(1+4\cosmo)}^{-n_h}
\paren{1+\mphi^2}^{-n_\phi}\paren{1+\mchi^2}^{-n_\chi}.
} 
The beta function of the scalar mass $\mphi^2$ is
\al{
{\tilde \beta}_{m_\phi^2}
&= \left(-2+\eta_{\phi} \right)\tilde{m}_{\phi}^2\nn
&\quad+
\frac{3\newton \mphi^2}{2\pi}
\Big[\beta^4 \left(24 \cosmo^2+16 \cosmo+1\right)+16 \beta ^3 \left(\cosmo^2-6
   \cosmo-1\right)+2 \beta ^2 \left(64 \cosmo^3-152 \cosmo^2+64
   \cosmo+43\right)
   \nn
&\qquad-48 \beta  \left(\cosmo^2-6 \cosmo+4\right) +312\cosmo^2-432 \cosmo+153\Big]\paren{1-\frac{\eta_h}{6}}
 {\mathcal I}\fn{2,2,0,0}\nn
&\quad
+\frac{12\newton \mphi^4}{\pi}\Big[(-3+\beta)^2\paren{1-\frac{\eta_h}{6}}{\mathcal I}\fn{0,2,1,0}
+\paren{1-\frac{\eta_\phi}{6}}{\mathcal I}\fn{0,1,2,0}
\Big]
\nn
&\quad-\frac{3\lphi}{32\pi^2}\paren{1-\frac{\eta_\chi}{6}}{\mathcal I}\fn{0,0,2,0}
-\frac{\lpchi}{64\pi^2}\paren{1-\frac{\eta_\phi}{6}}{\mathcal I}\fn{0,0,0,2}.
}
The beta function of the quartic coupling constant $\lphi^2$ is
\al{
{\tilde \beta}_{\lphi}
&=2\eta_{\phi}\tilde{\lambda}_{\phi}+\frac{3\newton\lphi}{2\pi} \Big[
\beta ^4 \left(24 \cosmo^2+16 \cosmo+1\right)
+16 \beta ^3
   \left(\cosmo^2-6 \cosmo-1\right)
   +2 \beta ^2 \left(64
   \cosmo^3-152 \cosmo^2
   +64 \cosmo+43\right)\nn
&\qquad
-48 \beta 
   \left(\cosmo^2-6 \cosmo+4\right)   +312 \cosmo^2-432 \cosmo+153
\Big]\paren{1-\frac{\eta_h}{6}}{\mathcal I}\fn{2,2,0,0}\nn
&\quad
+48\newton^2 \mphi^4 \Big[
3 \beta ^6 \left(32 \cosmo^3+32 \cosmo^2+4 \cosmo+1\right)+2 \beta ^5
   \left(32 \cosmo^3-288 \cosmo^2-96 \cosmo-19\right)
\nn
&\qquad
+5 \beta ^4 \left(-224
   \cosmo^3+96 \cosmo^2+180 \cosmo+49\right)
+12 \beta ^3 \left(32
   \cosmo^3+192 \cosmo^2-96 \cosmo-79\right)
   \nn
&\qquad
+3 \beta ^2 \left(736
 \cosmo^3-864 \cosmo^2-708 \cosmo+703\right)
+18 \beta  \left(32
  \cosmo^3-288 \cosmo^2+384 \cosmo-139\right)
  \nn
&\qquad
-9 \left(416
   \cosmo^3-864 \cosmo^2+612 \cosmo-147\right)
\Big]\paren{1-\frac{\eta_h}{6}}{\mathcal I}\fn{3,3,0,0}\nn
&\quad
-\frac{48 \newton \mphi^2 \lphi}{\pi}\Big[
(-3+\beta)^2 \paren{1-\frac{\eta_h}{6}}{\mathcal I}\fn{0,2,1,0}
+\paren{1-\frac{\eta_\phi}{6}}{\mathcal I}\fn{0,1,2,0}
\Big]\nn
&\quad
+768 \newton^2\mphi^6\Big[
2\paren{\beta ^4-6 \beta ^3+\beta ^2 (6-48 \cosmo)+18 \beta -27}\paren{1-\frac{\eta_h}{6}} {\mathcal I}\fn{0,3,1,0}
+ (-3+\beta^2) \paren{1-\frac{\eta_\phi}{6}}  {\mathcal I}\fn{0,2,2,0}
\Big]\nn
&\quad
+\frac{72\newton \mphi^4 \lphi}{\pi} \Big[
 (-3+\beta)^2  \paren{1-\frac{\eta_h}{6}}  {\mathcal I}\fn{0,2,2,0}
+2 \paren{1-\frac{\eta_\phi}{6}}  {\mathcal I}\fn{0,1,3,0}
\Big]\nn
&\quad
+9216\newton^2 \mphi^8\Big[ (-3+\beta)^2 \paren{1-\frac{\eta_h}{6}} {\mathcal I}\fn{0,3,2,0} 
+ \paren{1-\frac{\eta_\phi}{6}}{\mathcal I}\fn{0,2,3,0}
\Big]\nn
&\quad
+\frac{\lpchi^2}{64\pi^2}\paren{1-\frac{\eta_\phi}{6}} {\mathcal I}\fn{0,0,0,3}
+\frac{9\lphi^2}{16\pi^2}\paren{1-\frac{\eta_\chi}{6}} {\mathcal I}\fn{0,0,3,0}.
}
The beta function of the portal coupling constant $\lpchi$ is
\al{
{\tilde \beta}_{\lpchi}
&= \left(\eta_{\phi}+\eta_{\chi}\right)\tilde{\lambda}_{\phi\chi} + 
\frac{\newton \lpchi}{4\pi}
\Big[
\beta ^4 \left(24 \cosmo^2+16 \cosmo+1\right)
+16 \beta ^3
   \left(\cosmo^2-6 \cosmo-1\right)
+2 \beta ^2 \left(64
   \cosmo^3-152 \cosmo^2+64 \cosmo+43\right)\nn
   &\qquad
-48 \beta 
   \left(\cosmo^2-6 \cosmo+4\right)
+312 \cosmo^2-432\cosmo+153
\Big]\paren{1-\frac{\eta_h}{6}} {\mathcal I}\fn{2,2,0,0}\nn
&\quad
+\frac{96\newton^2\mphi^2\mchi^2}{\pi}\Big[
3 \beta ^6 \left(32 \cosmo^3+32 \cosmo^2+4 \cosmo+1\right)+2 \beta ^5 \left(32\cosmo^3-288 \cosmo^2-96 \cosmo-19\right)\nn
&\qquad
+5 \beta ^4 \left(-224\cosmo^3+96 \cosmo^2+180 \cosmo+49\right)
+12 \beta ^3 \left(32\cosmo^3+192 \cosmo^2-96 \cosmo-79\right)\nn
&\qquad
+3 \beta ^2 \left(736\cosmo^3-864 \cosmo^2-708 \cosmo+703\right)+18 \beta  \left(32\cosmo^3-288 \cosmo^2+384 \cosmo-139\right)\nn
&\qquad
-9 \left(416 \cosmo^3-864\cosmo^2+612 \cosmo-147\right)
\Big]\paren{1-\frac{\eta_h}{6}} {\mathcal I}\fn{3,3,0,0}\nn
&\quad
-\frac{24\newton \mphi^2 \lpchi }{\pi}\Big[
(-3+\beta)^2  \paren{1-\frac{\eta_h}{6}} {\mathcal I}\fn{0,2,1,0}
+  \paren{1-\frac{\eta_\phi}{6}} {\mathcal I}\fn{0,1,2,0}
\Big]\nn
&\quad
-\frac{24\newton \mchi^2 \lpchi }{\pi}\Big[
(-3+\beta)^2  \paren{1-\frac{\eta_h}{6}} {\mathcal I}\fn{0,2,0,1}
+ \paren{1-\frac{\eta_\chi}{6}} {\mathcal I}\fn{0,1,0,2}
\Big]\nn
&\quad+768 \newton^2 \mphi^4 \mchi^2 \Big[
2(\beta ^4-6 \beta ^3+\beta ^2 (6-48 \cosmo)+18 \beta -27) 
\paren{1-\frac{\eta_h}{6}} {\mathcal I}\fn{0,3,1,0}
\nn
&\qquad
+(-3+\beta^2) \paren{1-\frac{\eta_\phi}{6}} {\mathcal I}\fn{0,2,2,0}
\Big]\nn
&\quad
+768 \newton^2 \mchi^4 \mphi^2 \Big[
2(\beta ^4-6 \beta ^3+\beta ^2 (6-48 \cosmo)+18 \beta -27) 
\paren{1-\frac{\eta_h}{6}} {\mathcal I}\fn{0,3,0,1}
\nn
&\quad+(-3+\beta^2) \paren{1-\frac{\eta_\chi}{6}} {\mathcal I}\fn{0,2,0,2}
\Big]\nn
&\quad
+\frac{12\newton \mphi^4 \lpchi}{\pi}
\Big[
(-3+\beta^2)\paren{1-\frac{\eta_h}{6}} {\mathcal I}\fn{0,2,2,0}
+2\paren{1-\frac{\eta_\phi}{6}} {\mathcal I}\fn{0,1,3,0}
\Big]\nn
&\quad
+\frac{12\newton \mchi^4 \lpchi}{\pi}
\Big[
(-3+\beta^2)\paren{1-\frac{\eta_h}{6}} {\mathcal I}\fn{0,2,0,2}
+2\paren{1-\frac{\eta_\chi}{6}} {\mathcal I}\fn{0,1,0,3}
\Big]\nn
&\quad
+\frac{4\newton \mphi^2 \mchi^2 \lpchi}{\pi}
\Big[
(-3+\beta)^2\paren{1-\frac{\eta_h}{6}} {\mathcal I}\fn{0,2,1,1}
+\paren{1-\frac{\eta_\phi}{6}} {\mathcal I}\fn{0,1,2,1}
+\paren{1-\frac{\eta_\chi}{6}} {\mathcal I}\fn{0,1,1,2}
\Big]\nn
&\quad
+9216 \newton^2\mphi^4\mchi^4
\Big[
2(-3+\beta)^2 \paren{1-\frac{\eta_h}{6}} {\mathcal I}\fn{0,3,1,1}
+\paren{1-\frac{\eta_\phi}{6}} {\mathcal I}\fn{0,2,2,1}
+\paren{1-\frac{\eta_\chi}{6}} {\mathcal I}\fn{0,2,1,2}
\Big]
 \nn
&\quad
+\frac{\lpchi^2+\mphi^2\lpchi^2+3\lpchi \lphi +3\mchi^2\lpchi \lphi}{16\pi^2} \paren{1-\frac{\eta_\chi}{6}}  {\mathcal I}\fn{0,0,3,1}\nn
&\quad
+\frac{\lpchi^2+\mchi^2\lpchi^2+3\lpchi \lchi +3\mphi^2\lpchi \lchi}{16\pi^2} \paren{1-\frac{\eta_\phi}{6}}  {\mathcal I}\fn{0,0,1,3}.
}
The anomalous dimension $\eta_\phi$ is
\al{
\eta_\phi&= \frac{\newton}{6\pi} \Big[
3(-3+\beta)^2(-1+\beta)^2 \paren{1-\frac{\eta_h}{8}}{\mathcal I}\fn{0,2,1,0}
-(3+\beta)(-1+\beta)\paren{1-\frac{\eta_\phi}{8}}  {\mathcal I}\fn{0,1,2,0}
\Big].
}
\end{widetext}
Since the right-hand side involves the anomalous dimension $\eta_\phi$, we have to solve for $\eta_\phi$.
We see that the anomalous dimension vanishes for the choice $\beta=1$.
Note that there are no contributions from the matter coupling constants in symmetric phase.
The beta functions for $\mchi^2$, $\lchi$ and the anomalous dimension $\eta_\chi$ are obtained by the replacements $\mphi\leftrightarrow \mchi$, $\lphi \leftrightarrow \lchi$ and $\eta_\phi\leftrightarrow \eta_\chi$.
The anomalous dimension of the graviton $\eta_h$ is calculated, e.g., in  \cite{Christiansen:2012rx,Codello:2013fpa,Dona:2013qba,Christiansen:2014raa,Christiansen:2015rva,Dona:2015tnf,Denz:2016qks}. For models with few matter fields, such as the present one, $\eta_h$ is typically positive and smaller than one.
Note that although the anomalous dimension of the ghost field is neglected in the present work, it has been calculated in \cite{Eichhorn:2010tb,Groh:2010ta}.

\begin{figure*}
\begin{center}
\includegraphics[width=8cm]{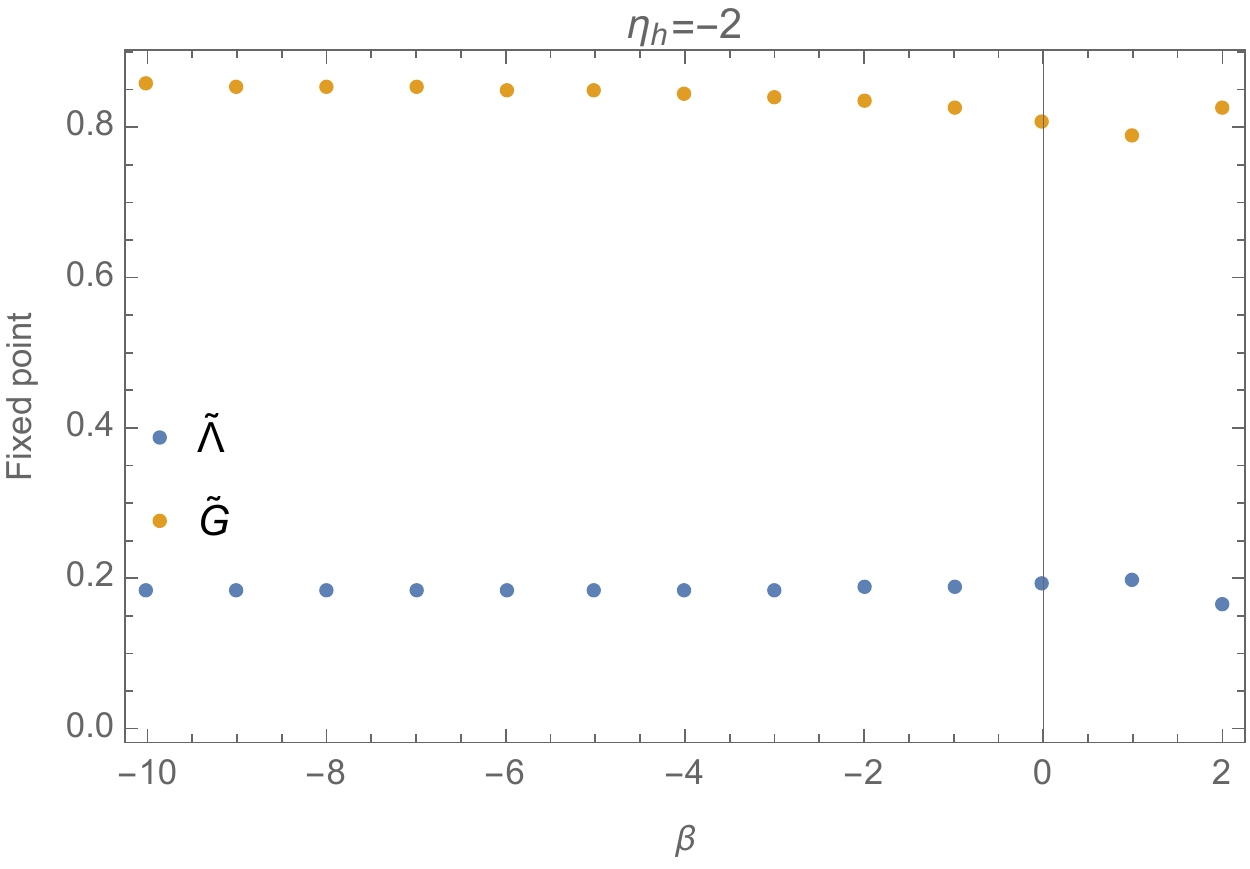}
\includegraphics[width=8.3cm]{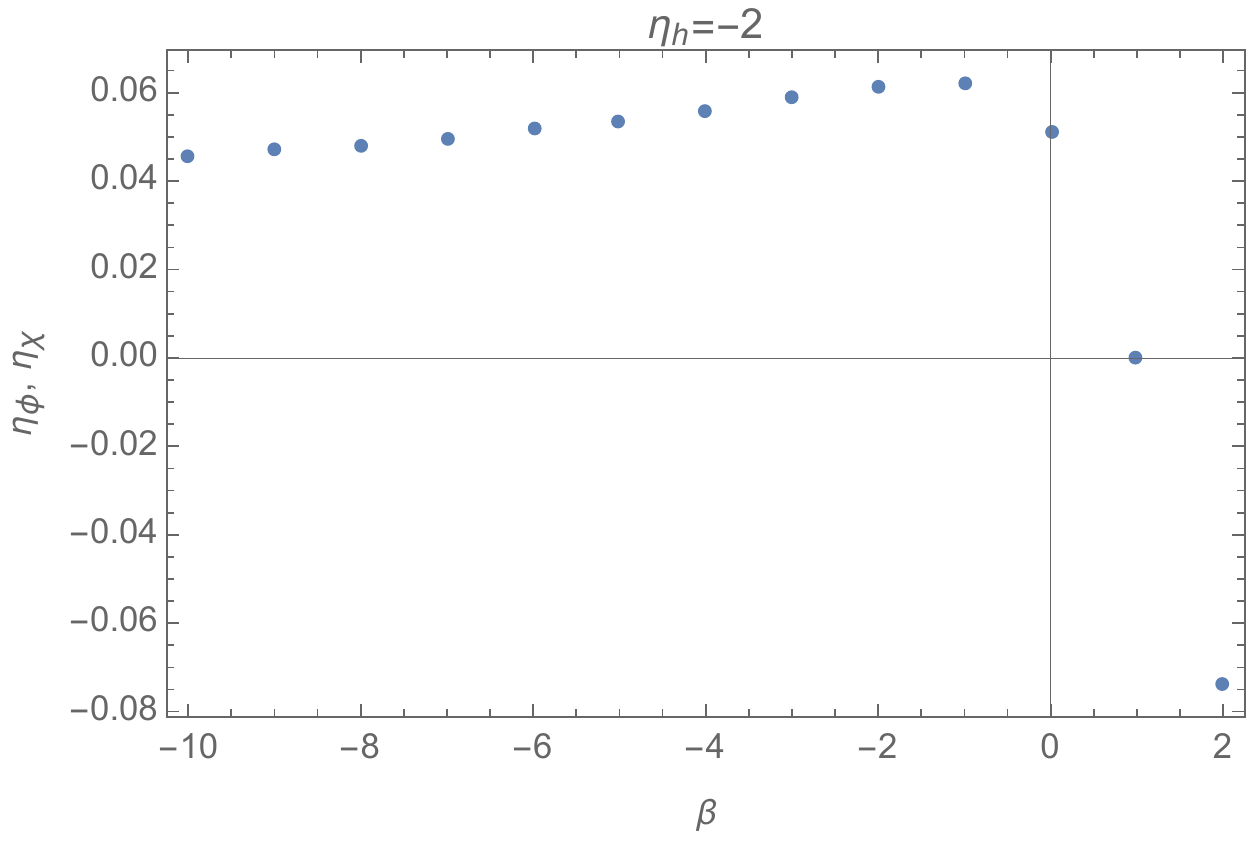}
\includegraphics[width=8cm]{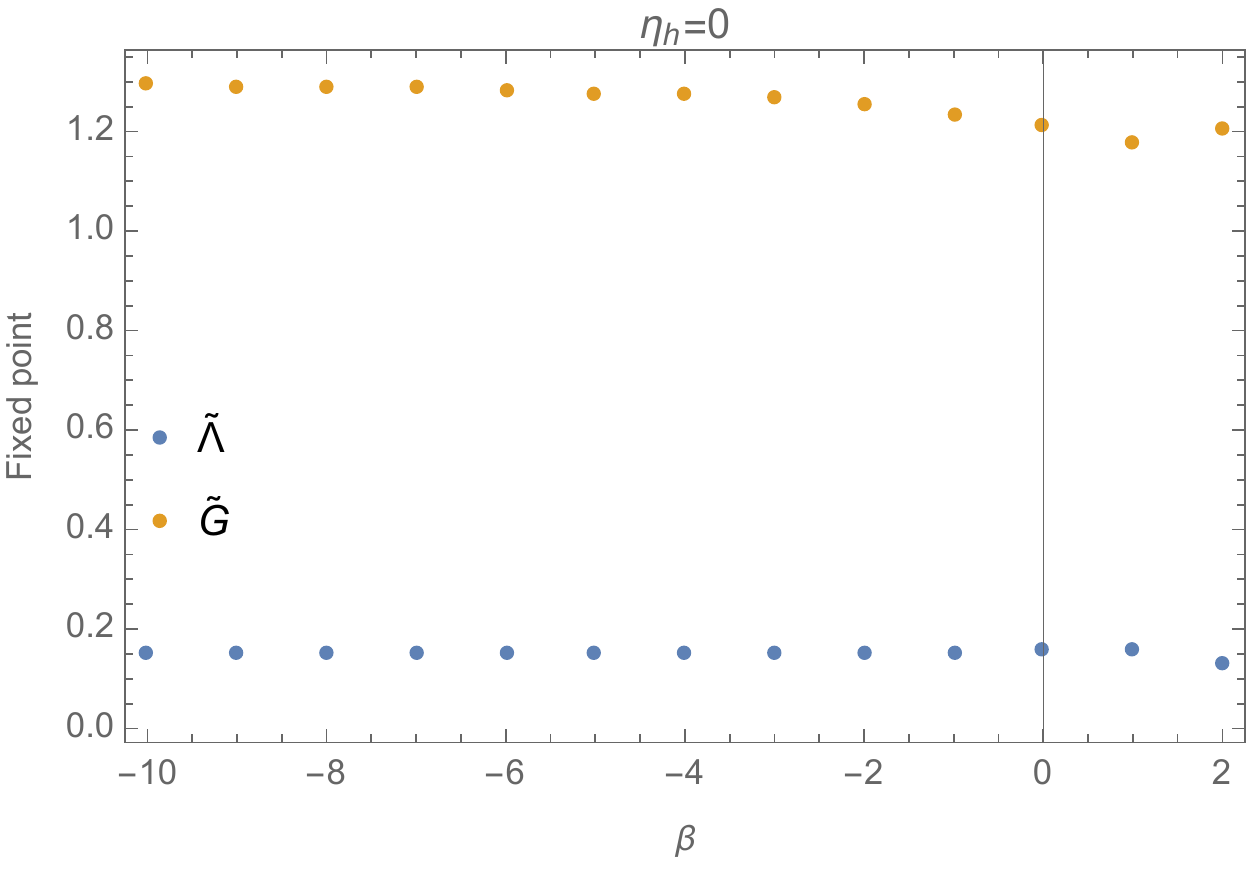}
\includegraphics[width=8.3cm]{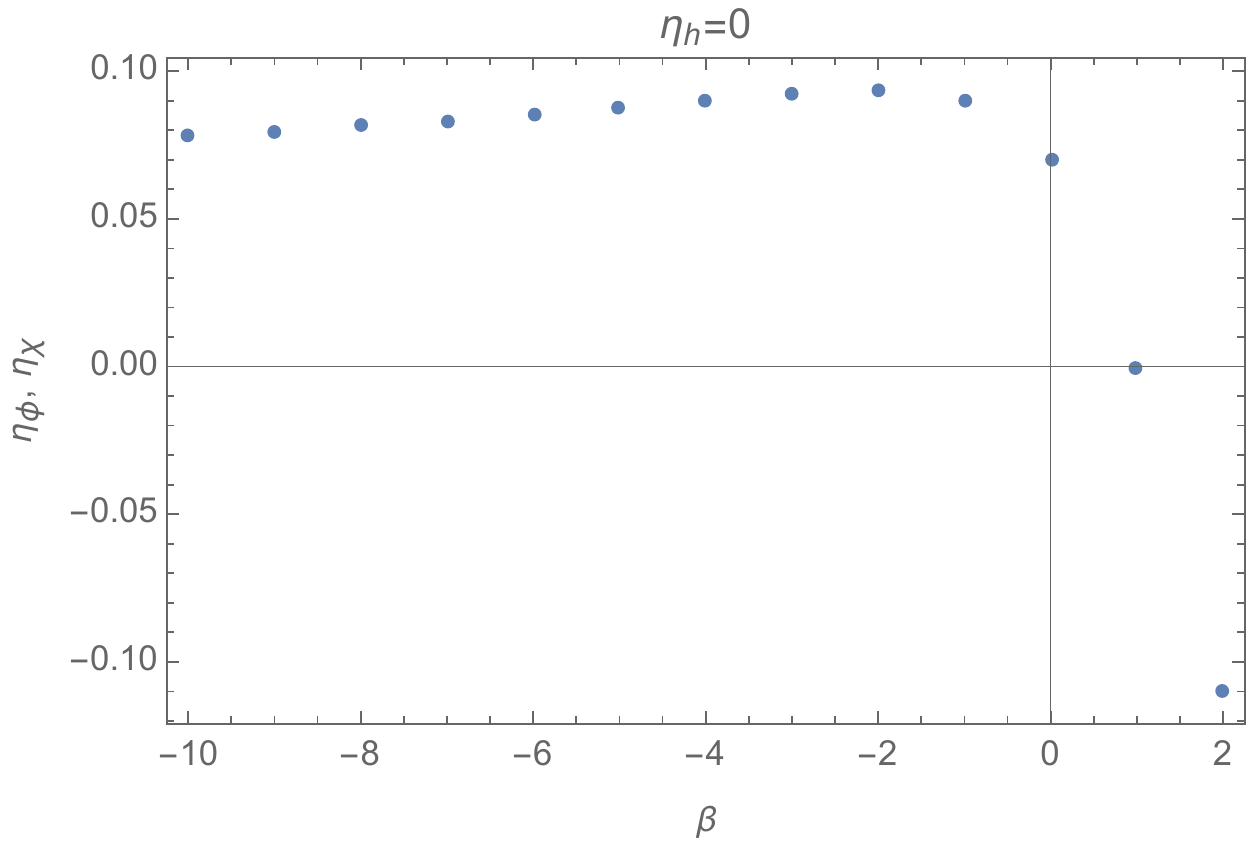}
\includegraphics[width=8cm]{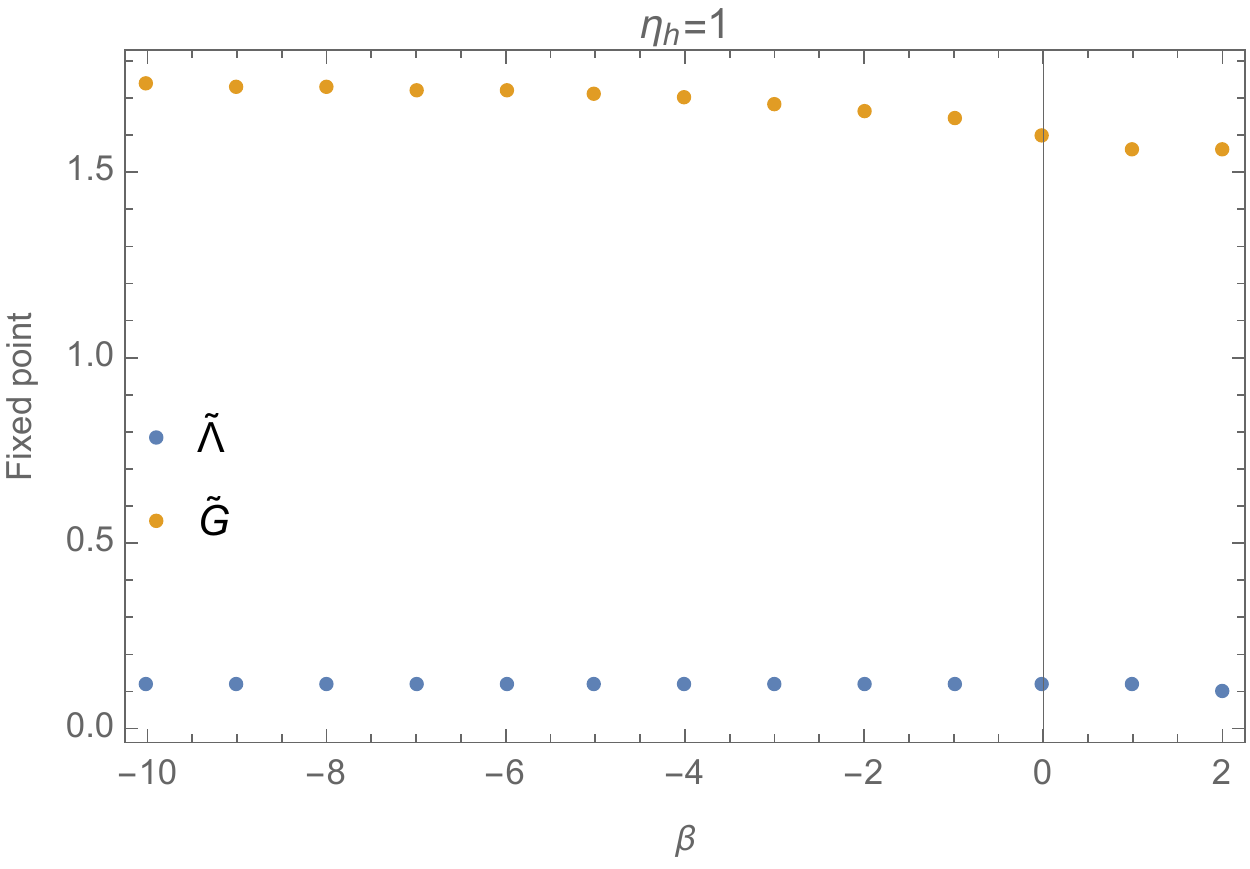}
\includegraphics[width=8.3cm]{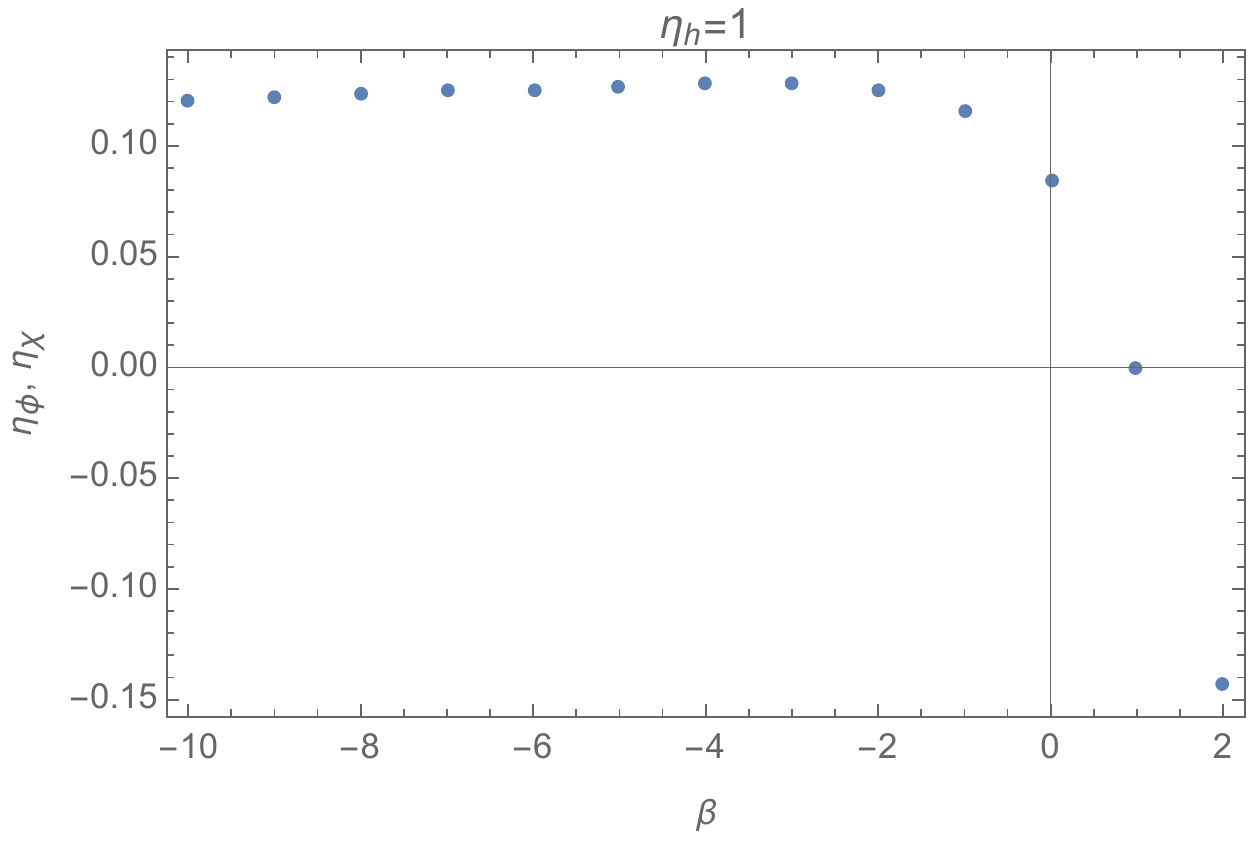}
\caption{Gauge dependence of the fixed point (left) and the anomalous dimensions (right) on the gauge parameter $\beta$ with $\eta_h=-2$, $0$ $1$. We choose $\alpha=0$.}
\label{values of fixed point and anomalous dimensions}
\end{center}
\end{figure*}

\begin{figure*}
\begin{center}
\includegraphics[width=12cm]{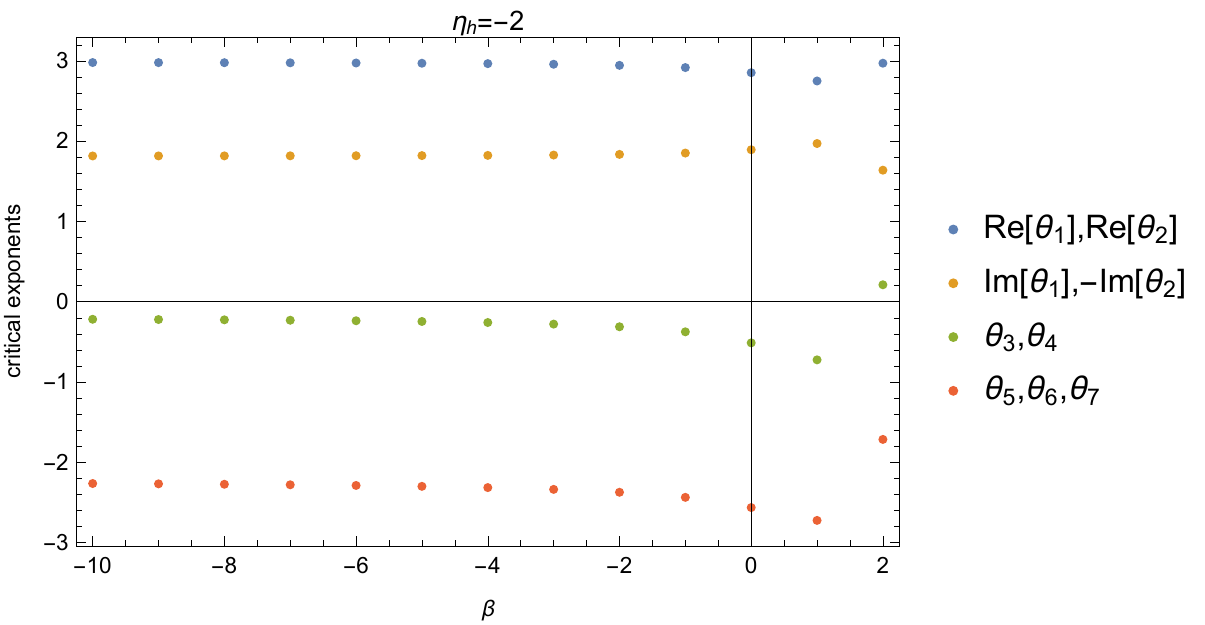}
\includegraphics[width=12cm]{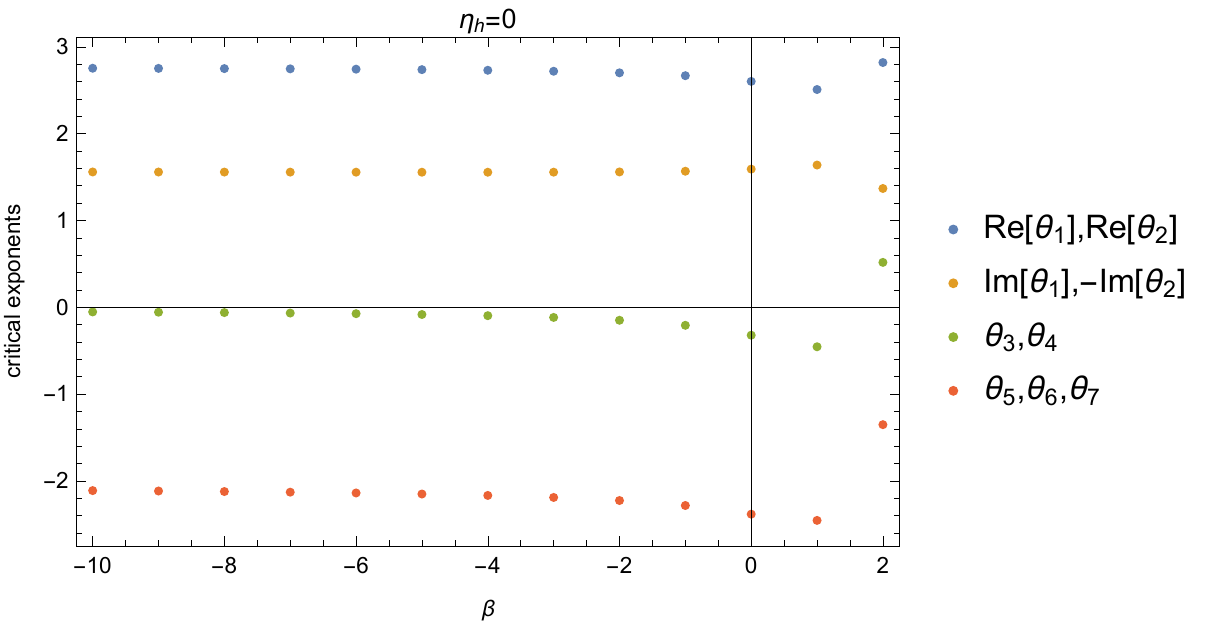}
\includegraphics[width=12cm]{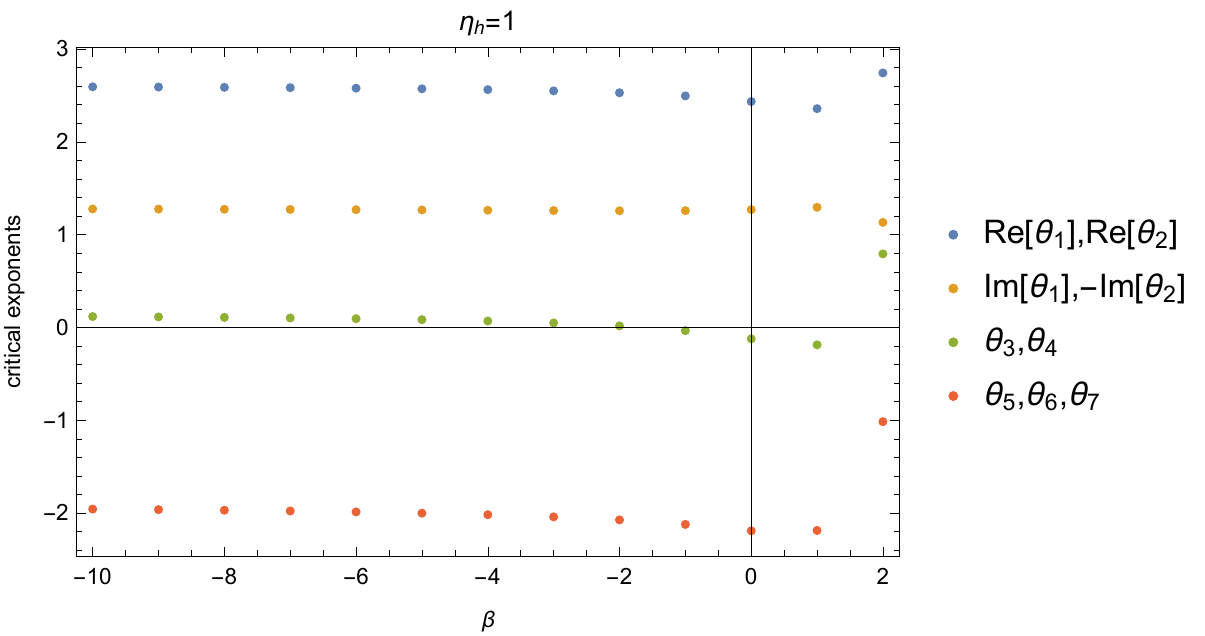}
\caption{Gauge dependence of the critical exponents on the gauge parameter $\beta$ with $\eta_h=-2$, $0$ $1$. We choose $\alpha=0$.}
\label{critical exponents}
\end{center}
\end{figure*}

\section{Gauge dependence}
We show the gauge dependences of the fixed point and critical exponents.
Here, we use $\alpha=0$, that is, the Landau gauge is employed.
Figure.~\ref{values of fixed point and anomalous dimensions} shows the dependence of the values of fixed point and the anomalous dimensions of scalar fields on $\beta$ with $\eta_h=-2$, $0$ and $1$.
For $\beta\leq0$, these values are stable under varying $\beta$.
The anomalous dimensions vanish at $\beta=1$.
The dependences of critical exponents on $\beta$ is shown in  Fig.~\ref{critical exponents}.
For $\eta_h=-2$, $0$, the critical exponents of scalar masses become negative.
In contrast, for $\eta_h=1$ their values depend on $\beta$.
For any case, the critical exponents of scalar masses with $\beta=2$ turn into positive.
This is because the beta functions have a pole at $\beta=3$.
Since the critical exponents are stable except for $\beta$ near the pole, we can conclude that only the Newton constant and the cosmological constant are relevant, while the matter coupling constants could be irrelevant for the smaller anomalous dimension of graviton field $\eta_h$.

\end{appendix}

\clearpage
\bibliography{refs}
\end{document}